\documentclass[twocolumn,showpacs,preprintnumbers,prd,floatfix]{revtex4} 

\usepackage[utf8]{inputenc}
\usepackage[dvipsnames]{xcolor}
\usepackage{subcaption}
\usepackage{amsmath} 
\usepackage{ifpdf,epsfig,floatflt,graphicx}
\DeclareGraphicsExtensions{.pdf, .jpg, .tif, .png}
\newcommand{\specialcell}[2][c]{%
  \begin{tabular}[#1]{@{}c@{}}#2\end{tabular}}
\begin{document}
\title[Self-gravitating stellar collapse]{Self-gravitating stellar collapse: explicit geodesics and path integration}
\author{Jayashree Balakrishna$^{1}$, Ruxandra Bondarescu$^{2}$, \\ Christine Corbett Moran*$^{3}$}
\address{$^{1}$College of Arts and Sciences, 317 HGA, Harris-Stowe State University, St. Louis, MO, USA \\ 
$^{2}$ University of Zurich, Zurich, Switzerland, CH-8057 \\
$^{3}$ NSF AAPF Fellow, TAPIR, California Institute of Technology, Pasadena, CA 91125\\
$^*$ corresponding author - corbett@tapir.caltech.edu}

\date{\today}

\begin{abstract}


We extend the work of Oppenheimer \& Synder to model the gravitational collapse of a star to a black hole by including quantum mechanical effects. We first derive closed-form solutions for classical paths followed by a particle on the surface of the collapsing star in Schwarzschild and Kruskal coordinates for space-like, time-like and light-like geodesics.  We next present an application of these paths to model the collapse of ultra-light dark matter particles, which necessitates incorporating quantum effects. To do so we treat a particle on the surface of the star as a wavepacket and integrate over all possible paths taken by the particle. The waveform is computed  in Schwarzschild coordinates and found to  exhibit  an ingoing and an outgoing component, where the former contains the probability of collapse,  while the latter contains the  probability that the star will  disperse.  These calculations pave the way for investigating the  possibility of quantum collapse  that does not lead  to black hole formation as well as for  exploring the nature of  the wavefunction inside $r=2M$. 

\end{abstract} 
\maketitle
\subsection{Introduction}


Black  holes play a  pivotal  role in  the  evolution of  the universe providing an  important test  laboratory  for general  relativity. 
Within classical general relativity, Oppenheimer \& Synder modelled the gravitational collapse of star to a black hole by approximating
the star with a uniform sphere of dust (hereafter O-S model) \cite{oppenheimer1939continued}. This model provides an analytic solution for stellar collapse 
that connects the Schwarzschild exterior of a star to a contracting Friedmann-Robertson-Walker (FRW) interior. Once the surface has passed within $r=2M$, no internal pressures can halt the collapse and all configurations collapse to a point-like singularity at $r=0$. The general features of this toy collapse model  have been  examined  by many authors \cite{goswami2004spherical,misner1973freeman,vaidya1951nonstatic,singh1997spherical,joshi2007gravitational}.

In the standard formalism from \cite{misner1973freeman}, the stellar surface is considered to be initially at rest. Here we consider configurations with all possible initial velocities. We derive closed-form solutions for the equations of motion in Schwarzschild and Kruskal coordinates for space-like, time-like and light-like geodesics. 

As an example application of our closed-form solutions for the classical O-S model with non-zero initial velocities, we consider the macroscopic collapse of a spherically symmetric sphere of dust composed of ultra-light particles. To approximate quantum effects, the radius of the star is approximated by a Gaussian wavepacket that is initially centred far from $r=2M$. Its evolution is then followed via a simple path integral approach that extends the results of Redmount and Suen \cite{1993IJMPA...8.1629R} from a relativistic free particle to a particle constrained by non-trivial gravity. Gravitational collapse incorporating approximate treatments of quantum mechanics has been considered by a variety of authors \cite{casadio1996semiclassical,hajivcek1992quantum,ortiz2007quantum,ansoldi1997classical,corichi2002quantum,alberghi1999gravitational,hawkins1994quantum,1998PhRvD..57.4812Z,Vaz:494112,Berezin:296102}. They involve non-equivalent ways of quantization that often produce physically different results \cite{dolgov1997properties}. Our path-integral approach approach is simpler, and can be easily compared to the assumption that the particle obeys the relativistic Schr\"{o}dinger equation.

In the classical model a star is idealised as a collapsing self-gravitating dust sphere of uniform density and zero pressure where the constituent particles have the attributes of classical dust: each particle  is assumed to  be infinitesimal in size  and to interact  only gravitationally with  other matter. The inclusion of quantum mechanical effects lifts  some of these assumptions allowing for  the possibility that some configurations will not  collapse to black holes but will disperse  or even form stable new configurations.
Quantitatively, quantum treatment is necessary in macroscopic stellar collapse when the action $S$ is of the order $\hbar$  \cite{narlikar1977quantum}.  In our case, the $S/\hbar \sim 1$ condition corresponds to $m M \sim \hbar$. For macroscopic black holes of masses $M = 1 M _\odot - 10^9 M_\odot$, this implies that quantum treatment is necessary for ultra-light constituent particles of $m \sim 10^{-10} - 10^{-19}$ eV.  In nature, such ultralight particle could be dark matter. Dark matter comes close to the attributes of classical dust, and some dark matter clouds may be dense enough to collapse to black holes. Particles as light as $10^{-22} - 10^{-23}$ eV have been proposed as constituents of dark matter halos \cite{lundgren2010lukewarm,1475-7516-2014-02-019,Hu2000,lesgourgues2002light}. 

To approximate quantum effects of the collapse of such a halo comprised of ultra-light particles, we start with an initial wavefunction that represents the position of the particle on the stellar
surface, and is at first far from $2 M$. To study the propagation of the wavefunction, we integrate over all possible paths taken by the particle.  At a given time, the outgoing wavefunction comprises the probability that the star disperses, and the ingoing wavefunction the probability that it collapses. We compute the propagator in analytical form for a particle on the surface of the star in the WKB approximation, and compare this to the limited assumption that the particle obeys the relativistic Schr\"{o}dinger equation. Our equations reduce to a free particle case when the mass of the star is zero; then the solution to the relativistic Schr\"{o}dinger equation is the exact representation of the wave function \cite{1993IJMPA...8.1629R}. In the more general case of a particle on the surface of a star,  this representation is no longer correct. However, the comparison is instructive. As expected, we observed that the WKB and relativistic Schr\"{o}dinger approximations are out of step at early and late times, and appear to converge towards one another at intermediate times.


In Schwarzschild coordinates, we can follow the evolution of the surface of the star only until the formation of an apparent horizon due to the $r=2M$ coordinate singularity. In Kruskal coordinates, one can continue to study the evolution of the star inside the apparent horizon ($r < 2 M$). The paths we have derived here include time-reversing space-like paths, which allow for the possibility of extraction of information from inside the horizon to the outside. These paths turn inside the horizon $r>0$, and head toward $r=2M$. In future work, explicit geodesic equations in Kruskal coordinates may be used as a stepping stone to model behaviour inside $r=2M$.

The rest of the paper is structured as follows: in \textbf{Section I} we describe the Oppenheimer-Snyder model. \textbf{Section II}  includes an overview of the classical action and paths, a computation of the Oppenheimer-Snyder limit and the derivation of the classical paths in Schwarzschild coordinates. \textbf{Section III} details the classical paths in Kruskal coordinates, which comprise a starting point for future work that explores the collapse inside $r=2M$.  As an example application of the closed form geodesics,  \textbf{Section IV}  describes the quantum treatment of the dust collapse applicable to ultra-light particles in Schwarzschild coordinates. The conclusions follow in \textbf{Section V}.
 
\section{The Oppenheimer-Snyder model}
 In General Relativity, a first approximation to the exterior space-time of any star, planet or black hole is a spherically symmetric 
 space-time modelled by the Schwarzschild metric. This is a consequence of Birkhoff's theorem \cite{misner1973freeman}.  
 The Schwarzschild line element thus takes the usual form
  \begin{equation}
ds^2 = - \left(1 - \frac{2 M}{r}\right) dt^2 + \frac{dr^2}{1 - 2 M/r} + r^2 d\Omega^2,
\end{equation}
where
\begin{equation}
d\Omega^2 = d\theta^2 + \sin^2 \theta d\phi^2.
\end{equation}
Throughout this paper we use geometric units with $G=c=1$.

 The Oppenheimer-Snyder (O-S) model follows the collapse of a star that is idealized as a dust sphere with uniform density 
 and zero pressure from the perspective of an observer located on the surface of the star. The motion of the collapsing surface initially at radius $r_i$ can be parametrized by 
\begin{eqnarray}
\label{OSeq1}
r &=& \frac{r_i}{2} \left(1 + \cos \eta \right) \\
\label{OSeq2}
\tau &=& \frac{r_i}{2} \sqrt{\frac{r_i}{2 M}} \left(\eta + \sin \eta \right) \\
\label{OSeq3}
t &=& 2 M \log\left|\frac{\sqrt{r_i/2 M -1} - \tan \eta/2}{\sqrt{r_i/2M -1} + \tan \eta/2}\right|  \\ \nonumber
& +& 2M \sqrt{\frac{r_i}{2 M} - 1} \left[\eta  
 + \frac{r_i}{4 M} (\eta + \sin \eta)\right] 
\end{eqnarray}
These equations correspond to Eqs. (32.10a) to (32.10c) of \cite{misner1973freeman}.
The star collapses to a singularity in finite proper time. However, it takes an infinite Schwarzschild $t$ to reach the apparent horizon at $r=2M$ and thus
an external observer will never see the star passing its gravitational radius  ($r=2M$).

In \cite{misner1973freeman} the stellar surface is considered to be initially at rest. Here we integrate the equations of motion
for configurations with all possible initial velocities.

\section{Classical Action and Radial Geodesics in Schwarzschild Coordinates}
\label{ClassicalEqs}
The relativistic action S and the Lagrangian ${\cal L}$ for this system are
\begin{equation}
S_{\rm cl} = - m \int d\tau = -m \int dt \sqrt{\left(1- \frac{2 M}{r}\right) - \frac{\dot{r}^2}{1 - 2 M/r}},
\label{OSaction}
\end{equation}
where $\tau$ is the proper time, and $\dot r = dr/dt$. The Lagrangian is
\begin{equation}
{\cal L} = - m  \sqrt{\left(1- \frac{2 M}{r}\right) - \frac{\dot{r}^2}{1 - 2 M/r}}
\label{OSLagrangian}
\end{equation}
The momentum
\begin{equation}
p = \frac{\partial {\cal L}}{\partial \dot r} = \frac{m \dot{r}}{\sqrt{(1- 2 M/r) \left[\left(1- 2 M/r \right)^2 - \dot{r}^2\right]}}
\end{equation}
and the Hamiltonian is 
\begin{equation}
H = p \dot{r} - {\cal L} = \sqrt{1 - \frac{2 M}{r}} \sqrt{m^2 + p^2 \left(1 - \frac{2 M}{r} \right)},
\label{Hameq}
\end{equation}
which reduces to the free particle Hamiltonian in the $M \to 0$ limit.
The equation of motion derived from the Euler-Lagrange equations is
\begin{equation}
- \frac{3 M}{r^2} \left(\frac{dr}{dt} \right)^2 + \left(1 - \frac{2 M}{r}\right) \frac{d^2 r}{dt^2} + \frac{M}{r^2} \left(1 - \frac{2 M}{r}\right)^2 =0,
\end{equation}
which can be integrated to 
\begin{equation}
\dot{r}^2 = \left(1- \frac{2 M}{r}\right)^2 \left[c_1 + \frac{2 M}{r} (1-c_1)\right].
\label{drdtsqr}
\end{equation}
From this the classical paths are determined
\begin{equation}
\int_{t_i}^{t_f} dt = t_f - t_i = \pm \int_{r_i}^{r_f} dr \frac{r^2}{x_r (r - 2M)},
\label{pathint}
\end{equation}
where 
\begin{equation}
x_r = \sqrt{c_1 r^2 + 2 M r (1-c_1)}.
\label{xr}
\end{equation}
The $+$ sign represents motions from $r_i$ to $r_f > r_i$, and $-$ sign represents the star collapsing from $r_i$ to $r_f < r_i$.
The parameter $c_1$ can be used to separate the space-time regions

\begin{eqnarray}
ds^2 > 0  \Longleftrightarrow  & c_1 > 1 \; \; \rm{ outside \; the \; light \; cone} \\  \nonumber
ds^2 = 0 \Longleftrightarrow  &  c_1 = 1  \; \; \rm{ on \; the \; light \; cone} \\ \nonumber
ds^2 < 0 \Longleftrightarrow  & c_1 < 1 \; \; \rm{ inside \; the \; light \;cone}\\
\end{eqnarray}

In the free particle case ($M \to 0$ in Eq.\ (\ref{drdtsqr})), $c_1=\dot{r}^2$ represents the velocity squared of the particle.
\begin{center}
{\bf Stellar Surface at Rest Limit}
\end{center}
 When the surface of the star is initially at rest, initial velocity $$\left|\frac{dr}{dt}\right|_{r=r_i} =0.$$ From Eqs. (\ref{drdtsqr}), this 
corresponds to
\begin{equation}
c_1 = - \frac{2 M}{r_i - 2 M}.
\label{c1zerospeed}
\end{equation}
We can recover our equations of motion for this $c_1$ value from the O-S model. We proceed by taking the derivative of Eq. (\ref{OSeq1}) and Eq. (\ref{OSeq3}) w.r.t. $\eta$
\begin{eqnarray}
\frac{dr}{d\eta} &=& - \frac{r_i}{2} \sin\eta  = - \sqrt{r (r_i - r)} \\ \nonumber
\frac{dt}{d\eta} &=&  \frac{r_i^2}{2 M}  \sqrt{\frac{r_i}{2 M} -1}  \left[\frac{\cos^4(\eta/2)}{(r_i/2 M) \cos^2 (\eta/2) -1}\right] \\ \nonumber
&=&\frac{r^2}{r - 2 M} \sqrt{\frac{r_i}{2 M} -1},
\end{eqnarray}
where we used that $r = r_i \cos^2 \eta/2$. We then divide $dr/d\eta$ by $dt/d\eta$, and use $r_i = 2M(c_1-1)/c_1$ from Eq.\ (\ref{c1zerospeed}) to recover the equation of motion
\begin{eqnarray}
\frac{dr}{dt}  = - \frac{r-2 M}{r^2}\sqrt{c_1 r^2 + 2 M r (1-c_1)}.
\end{eqnarray}
The negative sign $-$ fits in with our convention for a collapsing star.  While some paths will have low probability, a path integral approach that includes quantum effects requires
 the consideration of space-like, time-like and light-like paths with all possible initial velocities. 

\subsection{Classical Paths in Schwarzschild Coordinates}
\label{ClassicalPaths}
\begin{table*}
\centering
\resizebox{\textwidth}{!}{
\begin{tabular}{| l | l  |l |}
	\hline
  \textbf{Range of $\mathbf{r_i}$} & $\mathbf{c_1}$ & \textbf{Equations of Motion }\\
\hline
  $2 M-b_2$ & $c_1>0$ &  $t_f-t_i = \alpha_f - \alpha_i + M \frac{(3c_1 - 1)}{c_1\sqrt{c_1}}\log {\left(\frac{\beta_f}{\beta_i}\right)}$ \\
  $b_2-b_3$ & $-\frac{2M}{r_f-2M}<c_1<0$ & $t_f-t_i = \alpha_f - \alpha_i + \frac{M (3c_1 - 1)}{c_1 \sqrt{-c1}} \left[ \arcsin{\left(\frac{-c_1 r_f} {M (1 - c_1)} -1\right)} -  \arcsin{\left( \frac{-c_1 r_i}{M(1 - c1)}-1\right)} \right] $ \\
  $b_3-b_4$ & $-\frac{2M}{r_a-2M}$  & $t_f-t_i = 2\alpha_a  - \alpha_f -\alpha_i + M\frac{(3 c_1 - 1)}{c_1 \sqrt{-c_1}}\left[2 \arcsin{(1)} - \arcsin{\left(\frac{-c_1 r_f}{M(1 - c_1)} - 1\right)} - \arcsin{\left(\frac{-c_1 r_i}{M(1 - c_1)} -1 \right) }\right ]$ \\
  $b_4-b_5$ & $-\frac{2M}{b_4-2M}<c_1<0$ & $t_i-t_f = \alpha_f - \alpha_i +\frac{ M (3 c_1 - 1)}{ c_1 \sqrt{-c1}} \left[ \arcsin{\left(\frac{-c_1 r_f} {M (1 - c_1)} -1\right)} -  \arcsin{\left( \frac{-c_1 r_i}{M(1 - c1)}-1\right)} \right] $ \\
  $b_5- \infty $ & $c_1>0$ & $t_i-t_f = \alpha_f-\alpha_i +  M\frac{(3 c_1 - 1)}{c_1 \sqrt{c_1}}\log{ \left( \frac{\beta_f}{\beta_i} \right)}$ \\
\hline
\end{tabular}
}
\caption{Closed-form solution to Eq.\ (\ref{pathint}) in Schwarzschild coordinates.  At each region boundary $b_j$ ($j =2-5$) the equations converge to the form given in \textbf{Table \ref{jayatable1}}. All discontinuities cancel resulting in smooth transitions between regions without singularities.} 
\label{jayatable2}	
\end{table*}

\begin{table*}
\centering
\resizebox{\textwidth}{!}{
\begin{tabular}{ | l | l | l | }
	\hline
  \textbf{$\mathbf{r_i}$} &\textbf{$\mathbf{c_1}$} & \textbf{Regional Boundaries}\\
\hline
  $b_1$ & 1 & $t_f - t_i = r_f -b_1 + 2 M\log{\left(\frac{r_f - 2 M}{b_1 - 2M}\right)}$  \\
  $b_2$ & 0 & $t_f - t_i = \frac{\sqrt{2}}{3\sqrt{M}}\left(r_f^{3/2}-b_2^{3/2}\right) + 2\sqrt{2M}\left(\sqrt{r_f} - \sqrt{b_2}\right) +2M\log{\left[ \frac{\left(1 - \sqrt{\frac{2M}{r_f}}\right)\left(1 + \sqrt{\frac{2M}{b_2}}\right)}{\left(1 + \sqrt{\frac{2M}{r_f}}\right)\left(1 - \sqrt{\frac{2M}{b_2}}\right)}\right]}$ \\
  $b_3$ & $-\frac{2M}{r_f-2M}$  & $t_f-t_i = 2M \log{\left(1 - \frac{2M}{r_f}\right)}  -\frac{x_{b_3}}{c_1} +2 M \log{\left[\frac{(x_{b_3}+b_3)^2}{b_3 (b_3 - 2 M)}\right]} 
  + M \frac{(3 c_1 - 1)}{c1\sqrt{-c1}} \left[ \arcsin{(1)} - \arcsin{\left(\frac{-b_3 c_1}{M(1 - c_1)} - 1\right)}\right]$\\
   $b_4$ & $-\frac{2M}{b_4-2M}$ & $t_i-t_f =  \frac{x_f}{c_1}- 2 M\log{(1 - \frac{2M}{b_4})} - 2 M \log{\left[\frac{(x_f+r_f)^2}{r_f (r_f - 2 M)}\right]} 
+ M \frac{(3 c_1 - 1)}{c1\sqrt{-c1}} \left[-\arcsin{(1)} + \arcsin{\left(\frac{-r_f c_1 }{M(1 - c_1)} - 1\right)}\right]$ \\
  $b_5$ & 0 & $t_i - t_f = \frac{\sqrt{2}}{3\sqrt{M}}\left(r_f^{3/2}-b_5^{3/2}\right) + 2\sqrt{2M}\left(\sqrt{r_f} - \sqrt{b_5}\right) +2M\log{\left[\frac{\left(1 - \sqrt{\frac{2M}{r_f}}\right)\left(1 + \sqrt{\frac{2M}{b_5}}\right)}{\left(1 + \sqrt{\frac{2M}{r_f}}\right)\left(1 - \sqrt{\frac{2M}{b_5}}\right)}\right]}$ \\
  $b_6$ & 1 & $t_i - t_f = r_f -b_6 + 2 M\log{\left(\frac{r_f - 2 M}{b_6 - 2M}\right)}$ \\
\hline
\end{tabular}
}
\caption{Region boundaries for the equation of motion (Eq.\ (\ref{pathint})). The first column is the border value of $r_i$ for each region given the $r_f$ and $c_1$ values. It is calculated using the equation of motion in the third column. The light cone boundary corresponds to $c_1 =1$. }
	\label{jayatable1}
\end{table*}
Schwarzschild coordinates allow us to explore the space-time outside $r=2 M$.
The possible paths taken by a particle on the surface of a collapsing star are written explicitly in  \textbf{Table \ref{jayatable2}}.  Every point $(r_f, t_f)$ may be reached 
either via a direct path or via an indirect path. \textbf{Table \ref{jayatable1}} contains the boundary regions that determine
whether a path is direct or indirect, and whether it lies inside or outside the light cone. 

 The table uses
\begin{equation}
	x(r_s) = x_s=\sqrt{c_1 r_s^2 + 2Mr_s(1-c_1)},
	\label{Xrs}
\end{equation}
\begin{equation}
	\beta_s=\sqrt{c_1}x_s+c_1r_s+M(1-c_1),
	\label{beta}
\end{equation}
\begin{equation}
	\alpha_s=\frac{x_s}{c_1}+2M\log(r_s-2M) - 2M \log \left( \frac{(r_s+x_s)^2}{r_s}\right),
	\label{alpha}
\end{equation}
where $s = i,f,a$.    

The direct paths include a logarithmic term when $c_1 >0$ and an arc sin term when $c_1 < 0$. 
The indirect paths occur for $r_i$ in the $b_3 - b_4$ region. They each contain a 
zero velocity point where the path turns around. Mathematically, the velocity passes through zero ($dr/dt =0$ at some radius $r_a$) when 
$x(r_a)=0$, which corresponds to $c_1 = - 2M/ (r_a- 2 M)$. The indirect paths reach the final point $(r_f, t_f)$ after going through
a zero velocity point at $r_a (t_a < t_f)$ with $r_a > r_f$ and $r_a > r_i$. The paths from $r_i$ to $r_a$ correspond to the positive sign
 in Eq.\ (\ref{pathint}), while the path from $r_a$ to $r_f$ corresponds to the negative sign.


\begin{figure}
  \centering
  \caption{Schematic of paths.}
	\label{pathschematic}
	\begin{subfigure}{\linewidth}
	\caption{$r_f<r_i$}
	\includegraphics[width=0.8\linewidth]{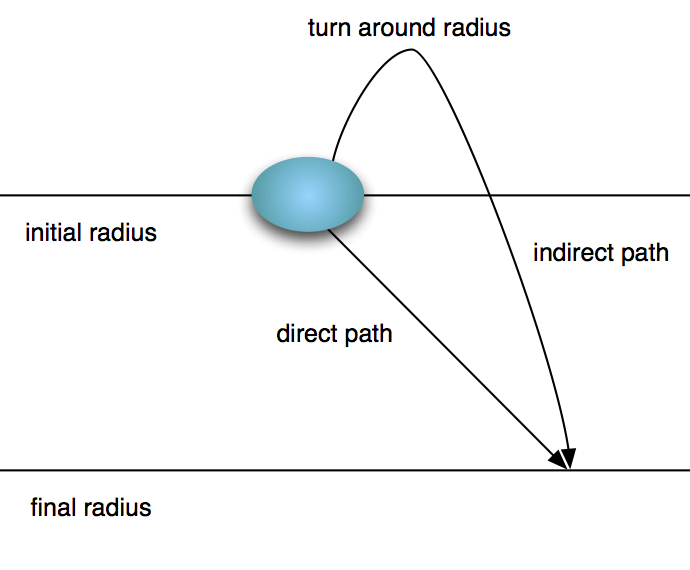}
	\end{subfigure}
	\begin{subfigure}{\linewidth}
	\centering
	\caption{$r_f=r_i$}
	\includegraphics[width=1.0\linewidth]{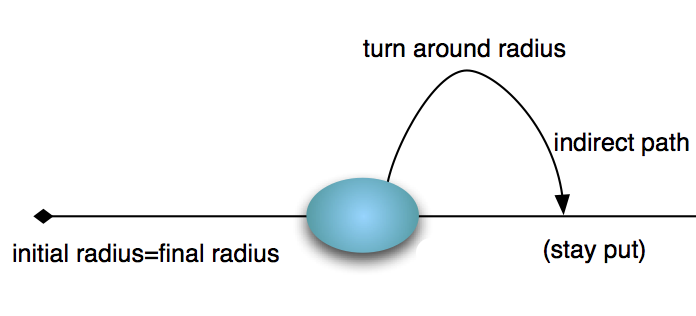}
	\end{subfigure}
      \begin{subfigure}{1.0\linewidth}
	\centering
	\caption{$r_f>r_i$}
	\includegraphics[width=0.8\linewidth]{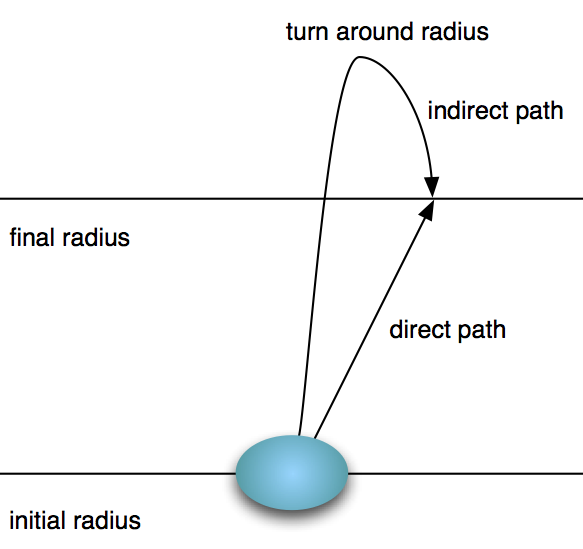}
	\end{subfigure}
\end{figure}

To understand these paths we make an analogy with the vertical motion of a ball under gravity (See Fig.\ \ref{pathschematic}).
This analogy arises naturally due to the correspondence between the Friedmann equation and the equation of motion 
for a test particle in the gravitational field of a point mass in Newtonian gravity.
It can be made only in the Schwarzschild case because then the interior of the star 
can be represented by an FRW model, which provides the link to Newtonian gravity.

Consider a ball traveling from point $r_i$ (initial radius) to point $r_f$ (final radius) with $r_f$ below $r_i$. It can either go down directly from $r_i$ to $r_f$ (direct path) or
it can go up from $r_i$, reach its highest point, where its velocity will be zero, and then fall back down to $r_f$ (indirect path). This is shown schematically in Fig. \ref{pathschematic}(a). Analogously, if a ball has to travel from $r_i$ to $r_f$ with $r_f$ above $r_i$,
 it can go directly from $r_i$ to $r_f$ or it can pass above $r_f$, reach its zero velocity point above $r_f$ and return down to $r_f$. This case can be seen in  Fig. \ref{pathschematic}(c). The middle figure (Fig. \ref{pathschematic}(b)) shows the special case when $r_i = r_f$. Then the ball can either stay put or it can be thrown up, reach its higher point, 
 and fall back down. Each of these paths takes a different amount of
time to complete. For a fixed travel time $\Delta t$, the two points $r_i$ and $r_f$ are connected by a single path that is either a direct path or a turn-around path. At the turning point of a turn-around path (also its highest point) the speed of the particle is always zero. 
  
If we keep the time interval constant, we find that for a given $r_f$ there is a point $r_i = b_4 > r_f$, where the velocity $dr/dt$ at $b_4 = 0$. 
Similarly, there is a point $r_i = b_3 <r_f$ such that the velocity $dr/dt$ at $r_f = 0$. For any point $r_i$ between $b_3$ and $b_4$, the only way a particle
can reach $r_f$ in the same time $\Delta t$ is if it  goes through a turning point $r_a > r_f$ and $r_a > r_i$.
The particle cannot reach $r_f > r_i$ if it's velocity vanishes at any point between $r_i$ and $r_f$.

 The motion towards or away from $r=2M$ can be compared to the vertical motion of a ball under gravity when thrown straight down or up, respectively.
 Classically, the star will always collapse to a black hole. Quantum mechanically, we include all possible initial velocities of the surface of the star towards and away from $2 M$. The wavefunction of a particle on the surface of the star determines the spread of the trajectory of the surface of the star itself. 

{\bf Escape Velocity.}
For a particle of mass $m$ that is thrown up from the surface of the Earth, an initial upward velocity $v_{\rm escape} \ge 11.2$ km/sec ensures that the particle escapes Earth gravity. At this speed 
\begin{equation}
    \frac{G M_E m}{R_E }= \frac{m v_i^2}{2},
\end{equation}
where $M_E$, and $R_E$ are the Earth mass and radius.

We continue with the ball analogy to estimate the escape velocity for our particle.  In our case, $R_E$ is replaced by the initial radius of the star $r_i$, and $v_i = dr/d\tau$. The kinetic energy of the particle on the surface of the star equals its potential energy when 
\begin{equation}
\frac{M m}{r_i} = \frac{m}{2} \left|\frac{dr}{d\tau}\right|_{r=r_i}^2.
\end{equation}
Here
\begin{equation}
\frac{dr}{d\tau} = \left(1- \frac{2 M}{r}\right)^{-1} \frac{dr}{dt}
\end{equation}
with $dr/dt = \dot{r}$ given by Eq.\ (\ref{drdtsqr}). This results in
\begin{equation}
\frac{M m}{r_i} = \frac{m c_1}{2} + \frac{M m}{r_i} (1 - c_1),
\end{equation}
whose solution is $c_1 = 0$. So, while all outgoing paths with $c_1 <0$ will turn-around, outgoing paths with $c_1 \ge 0$ will not
have turning points.  When $c_1 = 0$, the turning point ($dr/d\tau|_{r_a} = 0$,  $x_a =0$) occurs at $r_a \to \infty$ as expected for
 an escaped particle.


The different paths ({\bf Table \ref{jayatable2}}) and their various regional boundaries ({\bf Table \ref{jayatable1}})
are summarized below for a given $(r_f,t_f)$ outside $r=2M$:
\begin{itemize}
\item $\mathbf{c_1 = 1}$: the paths are light-like ($ds^2 = 0$) with $r_i = b_1  < r_f $ and $r_i = b_6 > r_f$. The paths with $\mathbf{c_1 > 1}$ are space-like, while those
with $\mathbf{c_1 <1}$ are time-like.
\item $\mathbf{c_1 = 0}$: $r_i = b_2 < r_f $  and $r_i = b_5 > r_f$. This path defines the escape velocity. All paths with $c_1 >0$ that are outgoing will not reverse, allowing the particle to escape the gravity
of the star.
\item $\mathbf{c_1 = - 2M/(r_f - 2 M)}$: $r_i = b_3 < r_f$. The minimum value of $r_i$ where $r_f$ is reached directly with $x(r_f) = 0$ and null final velocity  $$\frac{dr}{dt}|_{r = r_f}  = 0.$$ Paths with $r_i \le b_3$ are direct. Paths with $r_i$ between $b_3$ and $b_4$ are indirect.
\item $\mathbf{c_1 = - 2M/(b_4 - 2 M)}$: $r_i= b_4 > r_f$. This path starts with zero initial velocity. 
 All paths with $r_i > b_4$ are direct.
 \end{itemize}
\section{Classical Paths in Kruskal Coordinates}
 \label{Kruskal}
%
Our study is the first to consider space-like paths in Kruskal coordinates for a particle on the surface of a collapsing star. They have not been explicitly considered before because they are believed to be of low probability. We conjecture that they play an important role in quantum mechanical collapse in the same way the probability of passing through the potential barier is important in tunneling. Classically, it can never happen, and yet this probability cannot be ignored in quantum mechanics. 

In this section we derive and discuss the closed-form solutions to the equation of motion for the space-like, light-like and time-like paths in Kruskal coordinates. We find that space-like geodesics have the interesting property that they can turn around outside $r=2M$ and move back in time. Unlike the Schwarzschild $t$ and $r$, which are time and space coordinates respectively outside $r=2M$ but switch roles for $r<2 M$,  the Kruskal coordinate $v$ is always a time coordinate and the coordinate $u$ is always a space coordinate.  The relation between the Schwarzschild $r$ and $t$ and the Kruskal variables $u$ and $v$ is given by
\[ u = \left\{
  \begin{array}{l l}
    \sqrt{\frac{r}{2M} -1} e^{r/4M} \cosh{\frac{t}{4 M}} & \quad r > 2 M\\
   \sqrt{1-\frac{r}{2M}} e^{r/4M} \sinh{\frac{t}{4 M}} & \quad r < 2 M
  \end{array} \right.\]
  
  \[ v = \left\{
  \begin{array}{l l}
    \sqrt{\frac{r}{2M} -1} e^{r/4M} \sinh{\frac{t}{4 M}} & \quad r > 2 M\\
   \sqrt{1-\frac{r}{2M}} e^{r/4M} \cosh{\frac{t}{4 M}} & \quad r < 2 M
  \end{array} \right.\]

with 
\begin{equation}
u^2 - v^2 = \left(\frac{r}{2M} -1 \right) e^{r/2M}
\label{uvrelation}
\end{equation}
and
\begin{eqnarray}
\frac{v}{u} &=& \tanh \frac{t}{4M}, \;\; r > 2M \\ \nonumber
\frac{u}{v} &=& \tanh \frac{t}{4M}, \;\; r < 2M. 
\end{eqnarray}
Thus the line element in Kruskal coordinates takes the form
\begin{equation}
ds^2 = \frac{32 M^3}{r} e^{-r/2M} \left(du^2 - dv^2\right) + r^2 d\Omega^2.
\label{metricK}
\end{equation}
The singularity at $r=0$ occurs at $u^2 - v^2 = -1$. In this paper we consider only the first quadrant $u >0$ and $v>0$.  
 The horizon $r=2M$ occurs at $u = v$.  Outside the horizon $u > v$, whereas inside the horizon $u < v$. 

The classical equation of motion 
\begin{equation}
\frac{\ddot{u}}{1 - \dot{u}^2}  = \left(1- \frac{4 M^2}{r^2}\right) \frac{u - v \dot{u}}{u^2 - v^2}
\end{equation}
can be derived from the Euler-Lagrange equations
\begin{equation}
\frac{d}{dv} \frac{\partial {\cal L}}{\partial \dot{u}}  - \frac{\partial {\cal L}}{\partial u} =0,
\end{equation}
for the Lagrangian
\begin{equation}
{\cal L}(u,\dot{u},v) = - 4 M \sqrt{\frac{(1- \dot{u}^2) (1- 2 M/r)}{(u^2 - v^2)}}.
\end{equation}
Note that here $\dot{u} = du/dv$.

The equation of motion can be reduced to
\begin{equation}
\frac{du}{dv} = \frac{v/u \pm x_r/r}{1 \pm (v/u) (x_r/r)},
\label{dudv}
\end{equation}
where $x_r$ is given by Eq. (\ref{xr}). Direct paths from $r_i$ to $r_f> r_i$ have the ``$+$"-sign in the numerator and denominator with $du/dv > 0$.   Direct paths from $r_i$ to $r_f$ 
with $r_f<r_i$ will have the ``$-$"-sign in the numerator and denominator. 

The classical paths describing the evolution of the star outside $r = 2M$ have been discussed in Sec.\ \ref{ClassicalPaths} in Schwarzschild coordinates. It was shown that there were direct space-like and light-like paths as well as direct and indirect time-like paths. Any two points in $(r,t)$ were connected by a unique classical path.  

What is new in Kruskal space-time are the turning points
  in the $u-v$ plane for all space-like, light-like and time-like paths. Spacelike paths ($c_1>1$, $|du/dv| > 1$) turn in time at points when 
$dv/du = 0$, whereas timelike paths  ($c_1 < 1$, $|du/dv| < 1$) turn in space when $du/dv = 0$.  Additionally,
there are indirect space-like and light-like paths that have turning points in $r$ inside the apparent horizon (here
we refer to paths as being indirect when they turn in $r$; the light-like paths turn at $r=0$.)
  In contrast to the time-like paths,
the in-going space-like paths are {\it not} unique. A point outside $r = 2M$ might be connected to a point inside the apparent horizon via a direct and an indirect space-like path or via 
two indirect space-like paths. 
  
We determine the regions ({\bf Table  \ref{jayatableK2}}) and the light-cone boundaries ({\bf Table  \ref{jayatableK1}}) that
delimit the different kinds of paths that start at $(u_i, v_i =0)$ outside the horizon and reach $(u_f, v_f)$ inside the 
apparent horizon. Outside the horizon, the light cone boundary values 
of `$u_i$' for a given $u_f$ and $v_f$ can be determined by replacing $t_f - t_i$ in \textbf{ Table \ref{jayatable1}} by 
 $\tanh^{-1}(v_f/u_f)$.  The boundary values of $r_i$  ($v_i=0$) with $r_f<2M$, are calculated from the equations in {\bf Table  \ref{jayatableK1}}.
 Note that $u_f$ is a function of $v_f$  
 and $r_f$ and $u_i$ depends on $v_i$, which is typically taken to be zero, and $r_i$. Thus if one finds $r_i$, this determines $u_i$.  
The point $r_i=b_4'$ corresponds to $x_i=0$ with $c_1=-2M/(b_4'-2M)$. 
 The $c_1=0$ and $c_1=1$ (lightlike) paths that reach points $(u_f, v_f)$ inside the horizon are used to determine the $r_i$ values $b_5'$ and $b_{56}'$ respectively.
   The $r_i=b_6'$ value corresponds to $x_f=0$ with $c_1=-2M/(r_f-2M)$ (turning point at $r= r_f < 2 M$, space-like path, $c_1 > 1$). The value of $r_i=b_7'$ is to be determined by using $c_1=-2M/(r_a-2M)$  $(x_a=0, r_a<r_f<2M)$ in the equation of motion. The two unknowns $b_7'$ and $r_a$ are determined 
   by using the equation of motion and its partial derivative with respect to $c_1$. Both equations are given in \textbf{ Table \ref{jayatableK1}}. In the region outside $b_7'$ ($r_i >b_7'$),  there exist no classical paths to $(u_f,v_f)$.

\begin{table*}
\centering
\resizebox{\textwidth}{!}{
\begin{tabular}{| l | l  |l |}
	\hline
  \textbf{Range of $\mathbf{r_i}$} & $\mathbf{c_1}$ & \textbf{Equations of Motion for Paths with $r_i > 2M$ and $r_f < 2 M$.}\\
\hline
  $b_4'-b_5'$ & $-\frac{2M}{b_4'-2M}<c_1<0$  & $-4 M  \tanh^{-1} \left(\frac{u_f}{v_f}\right) = \alpha_f -\alpha_i + M\frac{(3 c_1 - 1)}{c_1 \sqrt{-c_1}}\left[\arcsin{\left(\frac{-c_1 r_f}{M(1 - c_1)} - 1\right)} - \arcsin{\left(\frac{-c_1 r_i}{M(1 - c_1)} -1 \right) }\right ]$ \\
  $b_5'-b_6'$ & $0<c_1<-\frac{2M}{r_f-2M}$ &$-4 M  \tanh^{-1} \left(\frac{u_f}{v_f}\right) = \alpha_f-\alpha_i +  M\frac{(3 c_1 - 1)}{c_1 \sqrt{c_1}}\log{ \left( \frac{\beta_f}{\beta_i} \right)} $ \\
  $b_{56}'- b_7' $ & $c_1=- \frac{2 M}{r_a-2 M}$ & $-4 M  \tanh^{-1} \left(\frac{u_f}{v_f}\right) = 2 \alpha_{a}- \alpha_f-\alpha_i +  M\frac{(3 c_1 - 1)}{c_1 \sqrt{c_1}}\log{ \left( \frac{\beta_{a}^2}{\beta_i \beta_f} \right)}$ \\
\hline
\end{tabular}
}
\caption{The equation of motion with $r_f < 2M$. At the light cone boundaries they converge to the form given in \textbf{Table \ref{jayatableK1}}. All discontinuities cancel each other and there are no singularities anywhere in the equations of motion.}
\label{jayatableK2}	
\end{table*}

\begin{table*}
\centering
\resizebox{\textwidth}{!}{
\begin{tabular}{ | l | l | l | }
	\hline
  \textbf{$\mathbf{r_i}$} &\textbf{$\mathbf{c_1}$} & \textbf{Light Cone Boundaries}\\
\hline
    $b_4'$ & $-\frac{2M}{b_4'-2M}$ 			& $-4 M  \tanh^{-1} \left(\frac{u_f}{v_f}\right) = \alpha_f - 2 M\log{\left|1 - \frac{2M}{b_4'}\right|} + M \frac{(3 c_1 - 1)}{c1\sqrt{-c1}} \left[-\arcsin{(1)} + \arcsin{\left(\frac{-r_f c_1 }{M(1 - c_1)} - 1\right)}\right]$ \\

	$b_5'$ & 		0 						& $-4 M  \tanh^{-1} \left(\frac{u_f}{v_f}\right) = \frac{\sqrt{2}}{3\sqrt{2 M}}\left(r_f^{3/2}-b_5'^{3/2}\right) + 2\sqrt{2M}\left(\sqrt{r_f} - \sqrt{b_5'}\right) +2M\log{\left[\frac{\left(\sqrt{\frac{2M}{r_f}}-1\right)\left(1 + \sqrt{\frac{2M}{b_5'}}\right)}{\left(1 + \sqrt{\frac{2M}{r_f}}\right)\left(1 - \sqrt{\frac{2M}{b_5'}}\right)}\right]}$ \\
  $b_{56}'$ & $1$ & $-4 M  \tanh^{-1} \left(\frac{u_f}{v_f}\right) = 2 M\log{\left|\frac{4M^2}{(r_f- 2 M)(b_{56}'-2M)}\right|} - r_f - b_{56}'$ \\

  $b_6'$ 	& $- \frac{2 M}{r_f - 2M}$		& $-4 M  \tanh^{-1} \left(\frac{u_f}{v_f}\right) = 2 M\log{\left(\frac{2M}{r_f}-1\right)} - \alpha_{b_6'} +  M \frac{(3 c_1 - 1)}{c1\sqrt{c1}} \log{\left(\frac{c_1 r_f + M (1-c_1)}{\beta_{b_6'}}\right)}$ \\

  $b_7'$ 	& $-\frac{2 M}{r_a-2M}$ 		& $-4 M  \tanh^{-1} \left(\frac{u_f}{v_f}\right) = 2 \alpha_a - \alpha_f - \alpha_{b7'} + M \frac{(3 c_1 - 1)}{c1\sqrt{c1}} \log{\left(\frac{\beta_{a}^2}{\beta_{b_7'} \beta_f}\right)} $, \\

  			&								& $\frac{x_{b_7'}^2 + 4 M b_7' (1-c_1)}{2 c_1^2 x_{b_7'}} + \frac{x_f^2 + 4 M r_f (1-c_1)}{2 c_1^2 x_f}
  - \frac{3 M (1-c_1)}{2 c_1^2 \sqrt{c_1}} \log\frac{\beta_{b_7'} \beta_f}{\beta_a} =0$ \\ \hline
\end{tabular}
}
\caption{Equations of motion at light cone boundaries for paths with $r_f < 2 M$. Outside $2M$ the equations 
for \textbf{Table \ref{jayatable2}} hold with $t_f - t_i$ replaced by $4 M \tanh^{-1}(v_f/u_f)$. 
Like before, the value $c_1=1$ corresponds to a light-like path. Note that  $b_7'$ is the caustic 
point where both the path $f(-2M/(r_a-2M), u, v)$ and its partial derivative $\partial f/\partial c_1$ vanish.}
	\label{jayatableK1}
\end{table*}

\begin{figure}
\begin{center}
\includegraphics[width=.8\linewidth]{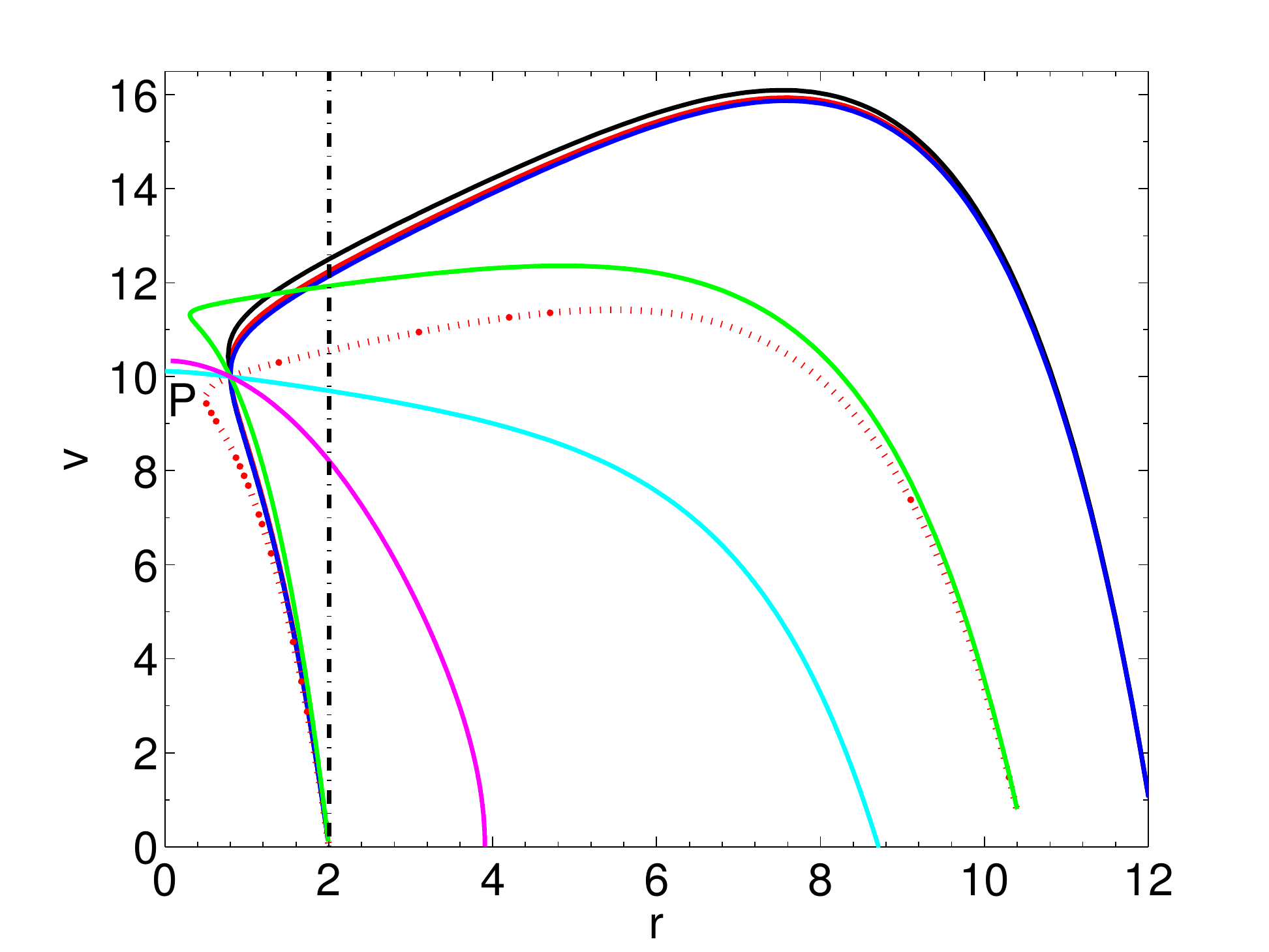}	
\includegraphics[width=.85\linewidth]{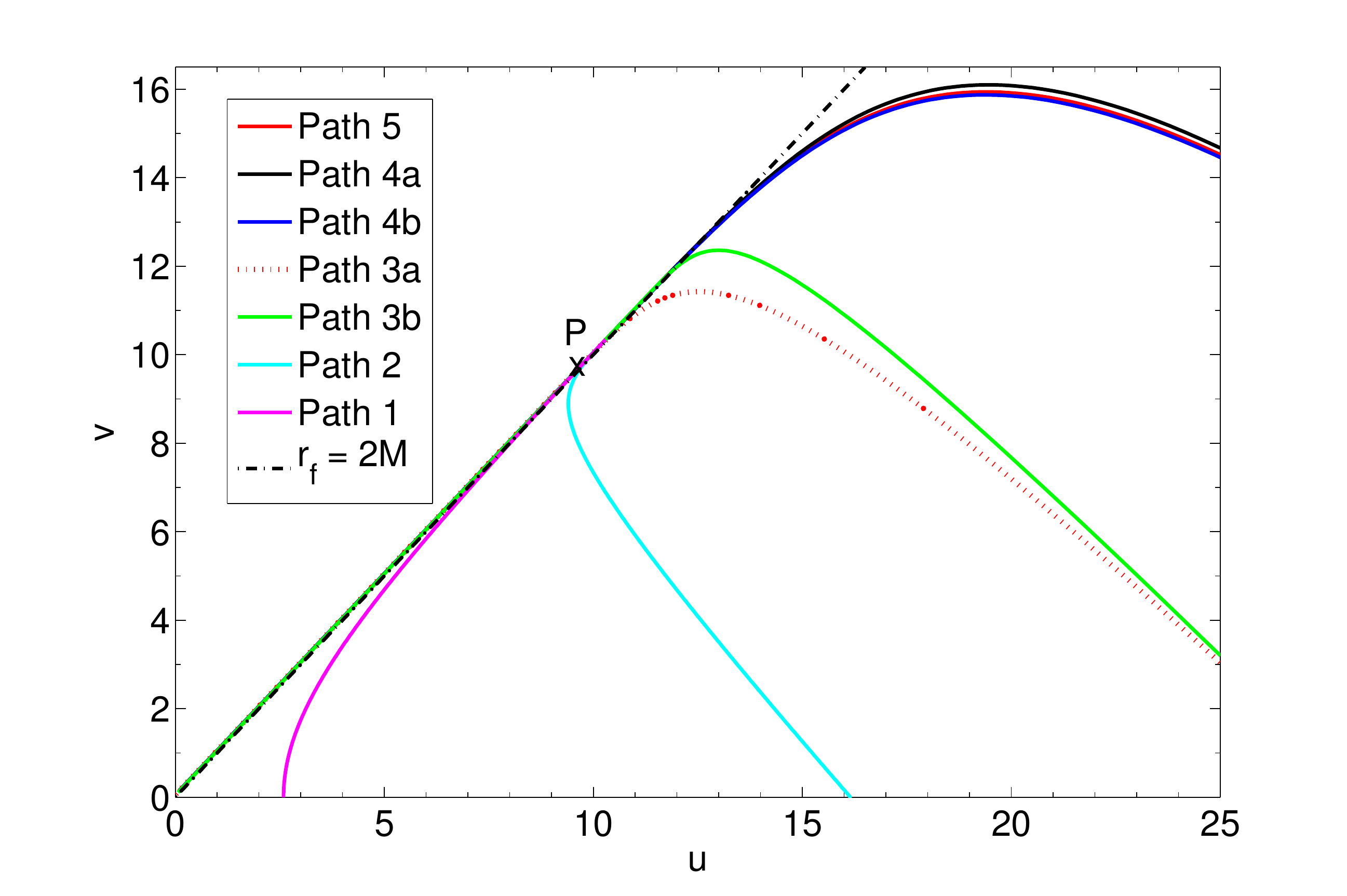}
\includegraphics[width=.8\linewidth]{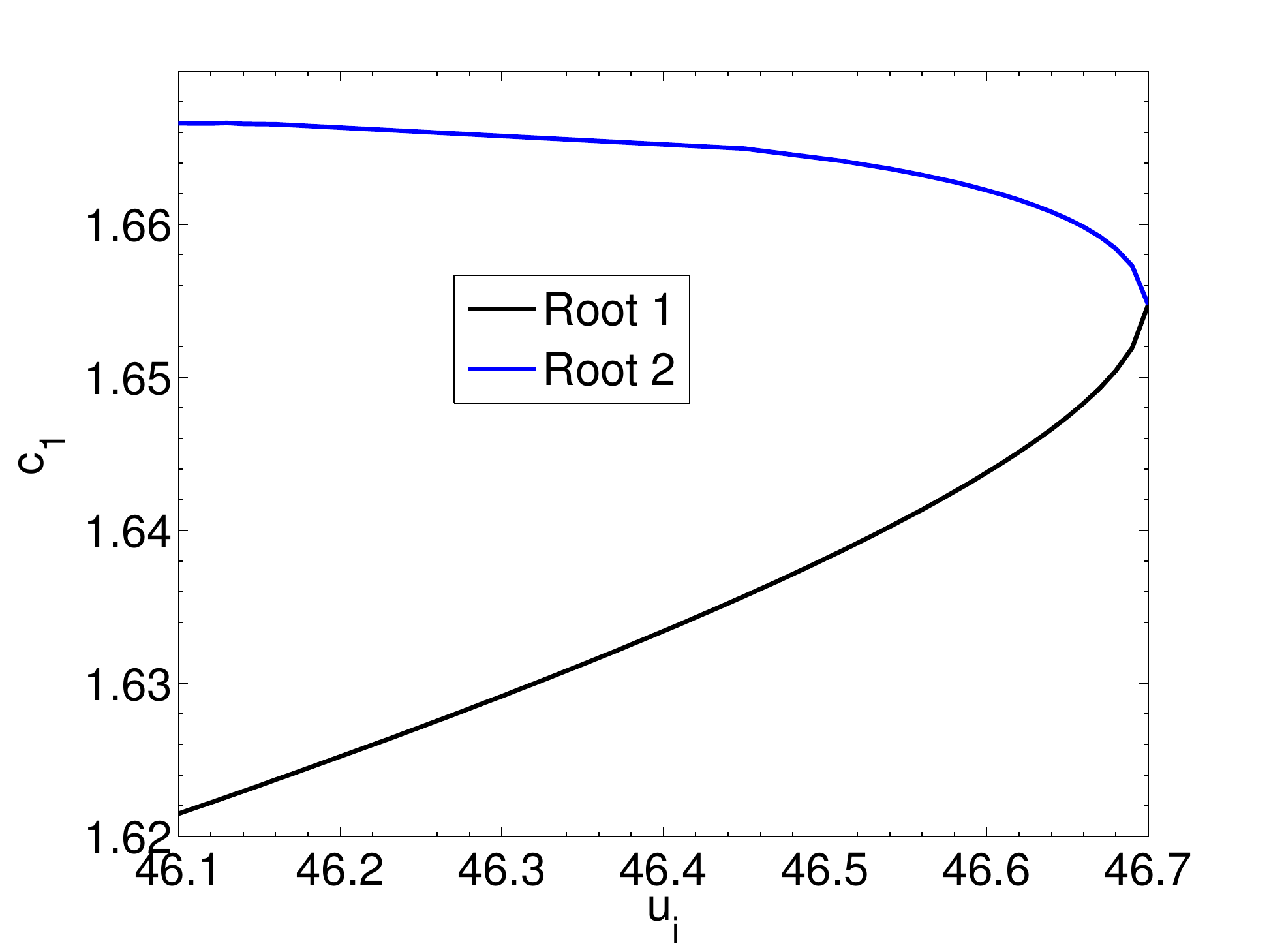}	
\end{center}
\caption{(a) $v$ vs $r$ and (b) $v$ vs $u$ are shown for space-like and time-like paths with $v_i=0$ that pass 
through $P(u_P = 9.995, v_P=10.0)$ of $r_P= 0.8$, a point located inside the apparent horizon.  (c) The $c_1$ values for
indirect space-like paths in the region between $b_{6}'$ and $b_7'$ from which point $P$ can be reached are displayed for each $u_i$. 
No paths exist beyond $r=b_7'$, which is where the two roots merge into one.}
\label{KruskalFig}
\end{figure} 
  In Fig. \ref{KruskalFig}(a) and (b) we trace direct time-like, direct space-like, and indirect space-like trajectories that start at $v_i = 0$
  and pass through  a fixed point  $\mathbf{P}(u_\mathbf{P} = 9.995, v_\mathbf{P}=10.0)$ with $r_\mathbf{P} \approx 0.8 M$, which is located inside the apparent horizon.
  {\bf Path 1} and {\bf path 2} are time-like, and originate at different points outside $r=2M$. {\bf Path 1} ($u_i = 2.59$, $c_1 = -1.05$) lies in the $b_4' - b_5'$ region of 
  {\bf Table  \ref{jayatableK2}}. {\bf Path 2} ($u_i = 16.15$, $c1 = 0.8$) 
  lies in the $b_5'-b_{56}'$ region, which contains time-like paths with a turning point in the $u-v$ plane ($du/dv =0$) 
  that lies outside the apparent horizon. 
  
  The next region is $b_{56}'- b_6'$, where each point outside the horizon is connected to points inside the horizon
  by two space-like paths: one direct and one indirect. {\bf Path 3a} and {\bf path 3b} originate at the same $u_i =  27.6$ ($r_i = 10.5 M$). 
  {\bf Path 3a} reaches $P$ directly. However, after having passed through P, it turns at $r_a = 0.48 M$ before reaching the horizon like all space-like paths.
  {\bf Path 3b} is indirect turning at $r_a = 0.3 M$ ($u_a = 11.3, v_a = 11.35$), and then reaching $P$ on its way back towards
  the apparent horizon. Both paths also have a $v-u$ turning point outside the apparent horizon where $dv/du = 0$. 
  After having passed through its $v-u$ turning point, each path moves back in time (the space coordinate  $u$ continuously decreases, while 
  the time coordinate $v$ increases up to the turning point and then decreases).
  
  In the $r_i = b_6' - b_7'$ region there are two indirect paths that connect the same point outside the horizon to a point inside
 $r=2 M$. {\bf Paths 4a} and {\bf 4b} connect the point of $u_i = 46.3$ to $P$. They each have a $dv/du =0$ turning point outside
 the horizon, and also turn in $r$ inside the horizon before reaching $P$.  {\bf Path 4a} turns at $r_a = 0.77147 < r_P$ and {\bf path 4b}
 turns at $r_a = 0.79969 < r_P$. It can be seen that the paths are very close together. As $u_i$ increases, the paths in this region become 
  progressively closer until they merge at $r_i = b_7'$. {\bf Path 5} shows the single indirect path that originates at $u_i = 46.7$ ($r_i = b_7' = 12.13 M$).
  Beyond this point, there are no real solutions and hence no way to reach point $P$. 
  
   Fig. \ref{KruskalFig}(c) displays the $c_1$ values of space-like paths reaching point $\mathbf{P}$ $(u_\mathbf{P} = 9.995, v_\mathbf{P}=10.0)$ from $u_i \ge 46.1$  at $v_i=0$. At first, for each $u_i$ in the figure there are two $c_1$ values with which the final point $\mathbf{P}$ can be reached. The figure clearly shows the roots ($c_1$ values) coming closer and closer together until they merge at the caustic point $r = b_7'$. Beyond this point there are no paths that reach point $\mathbf{P}$. Extensions to this work to determine the Kruskal-WKB wavefunction will require an analysis of the caustic at $r = b_7'$ to remove any potential divergences. 
   
   The indirect paths connecting $(u_i, v_i)$ with $r_i > 2 M$ to $(u_P, v_P)$ with $r_P < 2 M$ penetrate the horizon deeper than 
$r_P$ turning at $r_a < r_P$  before reaching $r_P$ (e.g., see the green path in Fig. \ref{KruskalFig}(a)).
After turning, all paths must continue towards the apparent horizon reaching it at $u=v=0$.  These paths move back
in time from a point outside the horizon, where $dv/du =0$, after having traveled to that point forward in Kruskal time from $(u_i, v_i=0)$
with $r_i > 2 M$.

As discussed before, the turning points in $r$ correspond to $x_r =0$ (in the Kruskal paths $x_r$ appears in Eq. (58)).
They occur at $r=r_a$ when $x_a =0$, where $c1=-2M/(r_a-2M)$. Since $c_1 >1$ for space-like paths, clearly $r_a < 2 M$
(the turning points are inside the horizon). For any $r$ on a path with $r_a$ as the turning point, $x_r$ can be written as
$\sqrt{2 M r (r- r_a)/(2 M- r_a)}$. To ensure the non-negativity of the term under the square root all points on this path
must have $r > r_a$. By the same token, for time-like paths $r_a > 2 M$ and $r < r_a$. This case was described in
the Schwarzschild analysis.
   
   All space-like paths that reach points inside the horizon subsequently head towards the horizon $r=2M$ ($u = v$). Classically, they end at $u_f=v_f =0$.
However,  quantum mechanically there may be paths close to the classical path that bring information from within the black hole horizon to the outside. 

\section{Quantum Treatment in Schwarzschild Coordinates}
\label{MainEqs}

\begin{table*}
\centering
\resizebox{\textwidth}{!}{
\begin{tabular}{ | l |l | l  | l | }
\hline
\specialcell{$c_1>0$ \\
            (direct path)}              & $\mathbf{S_{cl}}$                                               & $\pm i m \sqrt{c_1 -1} \left[ \frac{x_f-x_i}{c_1} - \frac{M (1-c_1)}{c_1 \sqrt{c_1}} \log\left( \frac{\beta_f}{\beta_i}\right) \right]$                                                                                           \\
                                        &                                                                  & \\
                                        & $\mathbf{\frac{\partial^2 S_{cl}}{\partial r_i \partial r_f }}$ & $ \frac{\mp i mr_i^2r_f^2 \left\{ \left(\frac{x_i+4Mr_i(1-c_1)}{2c_1^2 x_i}\right)-\left(\frac{x_f+4Mr_f(1-c_1)}{2c_1^2 x_f}\right)-\frac{3M(1-c_1)}{2c_1^{5/2}}\log{\left(\frac{\beta_i}{\beta_f}\right)} \right\}^{-1}}{2X_i X_f(r_i-2M)(r_f-2M)(c_1-1)^{3/2}} $   \\
                                        &                                                                  & \\
                                       \hline
\specialcell{$c_1<0$ \\ 
            (direct path)}              & $\mathbf{S_{cl}}$                                               & $\pm i m \sqrt{c_1 -1 } \left[ \frac{x_f-x_i}{c_1} - \frac{x_f-x_i}{c_1} \left(\arcsin\left( \frac{-c_1 r_f}{M(1-c_1)} -1 \right) - \arcsin\left(\frac{-c_1 r_i}{M(1-c_1)}-1 \right)\right)\right]$                             \\
                                        &                                                                  & \\
                                        & $\mathbf{\frac{\partial^2 S_{cl}}{\partial r_i \partial r_f }}$ &   $ \frac{\mp i mr_i^2r_f^2 \left\{ \left(\frac{x_i+4Mr_i(1-c_1)}{2c_1^2 x_i}\right)-\left(\frac{x_f+4Mr_f(1-c_1)}{2c_1^2 x_f}\right)+\frac{3M(1-c_1)}{2c_1^2 \sqrt{-c_1}} \left[ \arcsin{\left( \frac{c_1 r_f}{M(1-c1) } -1\right)} - \arcsin{\left( \frac{c_1 r_i}{M(1-c1) } -1\right)}\right] \right\}^{-1}}{2X_i X_f(r_i-2M)(r_f-2M)(c_1-1)^{3/2}}$   \\
                                        &                                                                  & \\

\hline
\specialcell{$c_1 = - \frac{2M}{r_a-2M}$ \\
            (indirect path)}            & $\mathbf{S_{cl}}$                                               & $\pm i m \sqrt{c_1-1} \left[ - \frac{x_f-x_i}{c_1} -\frac{M(1-c_1)}{c_1 \sqrt{c_1}} \left( 2 \arcsin{(1)} - \arcsin{\left( \frac{2r_i}{r_a}-1\right)} -  \arcsin{\left( \frac{2r_f}{r_a}-1\right)} \right)\right]$                  \\
                                        &                                                                  & \\
                                        & $\mathbf{\frac{\partial^2 S_{cl}}{\partial r_i \partial r_f }}$ &  $ \frac{+ i mr_i^2r_f^2 \left\{ \left(\frac{x_i+4Mr_i(1-c_1)}{2c_1^2 x_i}\right)-\left(\frac{x_f+4Mr_f(1-c_1)}{2c_1^2 x_f}\right)-\frac{3M(1-c_1)}{2c_1^2 \sqrt{-c_1}} \left[ \arcsin{\left( \frac{2 r_f}{r_a }-1\right)} - \arcsin{\left( \frac{2 r_i}{r_a } -1\right)} - 2 \arcsin{(1)}\right] \right\}^{-1}}{2X_i X_f(r_i-2M)(r_f-2M)(c_1-1)^{3/2}}$      \\
                                        &                                                                  & \\
                                       \hline
\end{tabular} 
}
\caption{The classical action in various regions. The $+$ sign is chosen when $r_f > r_i$, and the $-$ sign 
when $r_f < r_i$.} 
  \label{jayatable3}
\end{table*}
We consider a collapsing cloud of ultra-light particles with $m M \sim \hbar$. Such particles are not localized and thus 
neither is the surface of the star. The position of a particle on the surface of the star is then approximated by a wavefunction.  In a relativistic path
 integral approach, this wavefunction can be computed in the WKB approximation via an integral over the classical action \cite{Schulman:2012ud}. It is necessary to include configurations
  will all possible initial velocities. For each point $(r_f, t_f)$ all classical paths derived above are needed. In this paper, we only perform the WKB analysis in Schwarzschild 
  coordinates, where we can only use the paths to $r_f > 2M$.

\subsection{WKB Approximation: Closed-form Propagator, Numerical Wavefunction}
\label{WKBSubSection}

In order to study the wavefunction we use a WKB approximation of the propagator
\begin{equation}
G(r, t; r_i, t_i) = \int_C {\cal D}r \exp[i S/\hbar],
\end{equation}
where $S$ is the action associated with each path. The set of paths $C$ include space-like, time-like and light-like paths.
 
The WKB approximation involves expanding about the classical action $S_{\rm cl}$ to include paths that slightly deviate from 
the classical paths. The propagator is dominated by paths near the classical trajectory between $(r_i, t_i)$ and $(r,t)$ and is approximated
by the WKB expression \cite{Schulman:2012ud} 
\begin{equation}
G_{\rm WKB}(r ,t ;r_i,t_i) = \sqrt{\frac{i}{2 \pi \hbar} \frac{\partial^2 S_{\rm cl}}{\partial r_i \partial r_f}} \exp\left(i \frac{S_{\rm cl}}{\hbar} \right).
\label{GWKB}
\end{equation}
The WKB wave function is obtained from the integration of the propagator
\begin{equation}
\Psi_{\rm WKB}(r_f,t_f) = \int_{0}^{\infty} G_{\rm WKB}(r_f,t_f;r_i,t_i) \Psi(r_i,t_i) d r_i.
\label{WKBPsi}
\end{equation}

 The initial wavefunction that describes a particle on the surface of a star of radius $r_i$ is taken to be a Gaussian centered about $r_c$
\begin{equation}
\Psi(r_i,t_i=0) = \exp{\left[- \frac{\left(r_i - r_c \right)^2 m^2}{\hbar^2}\right]}
\label{Psi0}
\end{equation}
with $r_c  >> 2 M$.

The action from Eq.\ (\ref{OSaction}) is rewritten using Eqs. (\ref{drdtsqr}) and (\ref{xr}) as
\begin{equation}
S_{\rm cl}(r_i, t_i;r_f,t_f) = \pm i m \sqrt{c_1 - 1} \int_{(r_i, t_i)}^{(r_f, t_f)} dr \frac{r}{x(r)}
\end{equation}
for direct paths connecting $(r_i,t_i)$ to $(r_f,t_f)$. The $+$ sign corresponds to $r_f  > r_i$ and the $-$ sign corresponds
to $r_f < r_i$.

Indirect paths connecting $(r_i, t_i)$ to $(r_f, t_f)$ through the turning point $(r_a, t_a)$ with $r_a > r_i$ and $r_a > r_f$ are described
by the classical action
\begin{equation}
S_{\rm cl} = i m \sqrt{c_1 - 1} \left[ \int_{(r_i, t_i)}^{(r_a, t_a)} dr \frac{r}{x(r)} - \int_{(r_a, t_a)}^{(r_f, t_f)} dr \frac{r}{x(r)} \right].
\end{equation}

In Schwarzschild coordinates, the wavefunction is continuous across all regions outside $r= 2M$. The presence of the $r=2M$ coordinate
singularity prohibits the inclusion of paths at or beyond the horizon region. Closed-form expressions for $S_{\rm cl}$ and $\partial^2 S_{\rm cl}/(\partial r_i \partial r_f)$ are given in \textbf{Table \ref{jayatable3}}. Both functions are continuous across all regions outside $r=2M$. Once $c_1$ is known for every point $(r, t)$, we have all the ingredients to find the propagator in Eq. (\ref{GWKB}) using the 
 $S_{\rm cl}$, and $\partial^2 S_{\rm cl}/\partial r_i \partial r_f$ expressions from \textbf{Table \ref{jayatable3}}.
 The propagator is then integrated to find the wavefunction.

\subsection{Relativistic Schr\"{o}dinger Solution}
\subsubsection{Free Particle}
\label{FreeParticleSection}
The $M \to 0$ limit describes a relativistic free particle that is no longer confined by the gravity of the sphere of dust.
This case was first studied by Redmount and Suen in \cite{1993IJMPA...8.1629R}. We revisit this limit to investigate whether
 the wavefunction in the WKB approximation converges to the relativistic Schr\"odinger solution, which is the exact solution
for this case. In \cite{1993IJMPA...8.1629R} some disagreement was observed, but while they attempted a physical interpretation, it is likely due to numerical error.

When $M \to 0$, the Hamiltonian from (\ref{Hameq}) reduces to
\begin{eqnarray}
H = \sqrt{{\bf p}^2 + m^2},
\end{eqnarray}
where ${\bf p}$ is now the momentum operator. The Hamiltonian is nonpolynomial in ${\bf p}$. The square root term thus
corresponds to a non-local momentum operator that is interpreted as acting on any wavefunction  \cite{1993IJMPA...8.1629R}
\begin{equation}
\Psi(x,t) = \int_{-\infty}^{\infty} dk e^{i k x} \phi(k,t).
\end{equation}
to give
\begin{equation}
H \Psi(x,t) = \int_{-\infty}^{\infty} dk e^{i k x} \sqrt{\hbar^2 k^2 + m^2} \phi(k,t).
\end{equation}
This integrates to
\begin{equation}
\phi(k,t) = A(k,x_0) \exp\left(- \frac{i \Delta t}{\hbar} \sqrt{\hbar^2 k^2 + m^2}\right).
\end{equation}
When $\Delta t =0$, we recover the original wavefunction, which is chosen to be a Gaussian centred around the origin,
 $$\Psi(x_0,t_0) = {\cal N} \exp\left(-\frac{m^2 x_0^2}{\hbar^2} \right),$$ and so
\begin{eqnarray}
\Psi(x,t) &=& \frac{{\cal N}}{2 \pi} \int_{-\infty}^{\infty} dk \int_{-\infty}^{\infty} dx_0 \exp\left(-m^2 x_0^2/\hbar^2\right) \\ \nonumber
&\times& \exp\left(i k \Delta x\right) \exp\left(- \frac{i \Delta t}{\hbar} \sqrt{\hbar^2 k^2 + m^2}\right).
\label{psifreeschro}
\end{eqnarray}
The normalization $|{\cal N}|^2 = m \sqrt{2}/(\hbar \sqrt{\pi})$ is time independent.

In terms of the propagator
\begin{equation}
\Psi(x_f,t_f) = \int_{-\infty}^\infty d x_0 G(x_f, t_f; x_0, t_0) \Psi(x_0,t_0),
\label{PsiExact}
\end{equation}
where
\begin{eqnarray} \nonumber
G(x,t; x_0, t_0) &=& \int_{-\infty}^{\infty} \frac{dk}{2 \pi} \exp\left(i k \Delta x\right) \\ \nonumber
&&\times \exp\left[- \frac{i \Delta t}{\hbar} \sqrt{\hbar^2 k^2 +m^2}\right], \nonumber \\
&=& \lim_{\epsilon \to 0^+} \frac{m (i \Delta t + \epsilon)}{\pi \hbar \sqrt{\lambda_\epsilon}} K_1(m \lambda_\epsilon^{1/2}/\hbar).
\end{eqnarray}

Here $K_1$ is the modified Bessel function and $\lambda_\epsilon \equiv \Delta x^2 + (i \Delta t + \epsilon)^2$.  When $\Delta t = 0$ the propagator reduces to $\delta(\Delta x)$ as expected. 
\begin{figure}\begin{center}
\includegraphics[width=.85\linewidth]{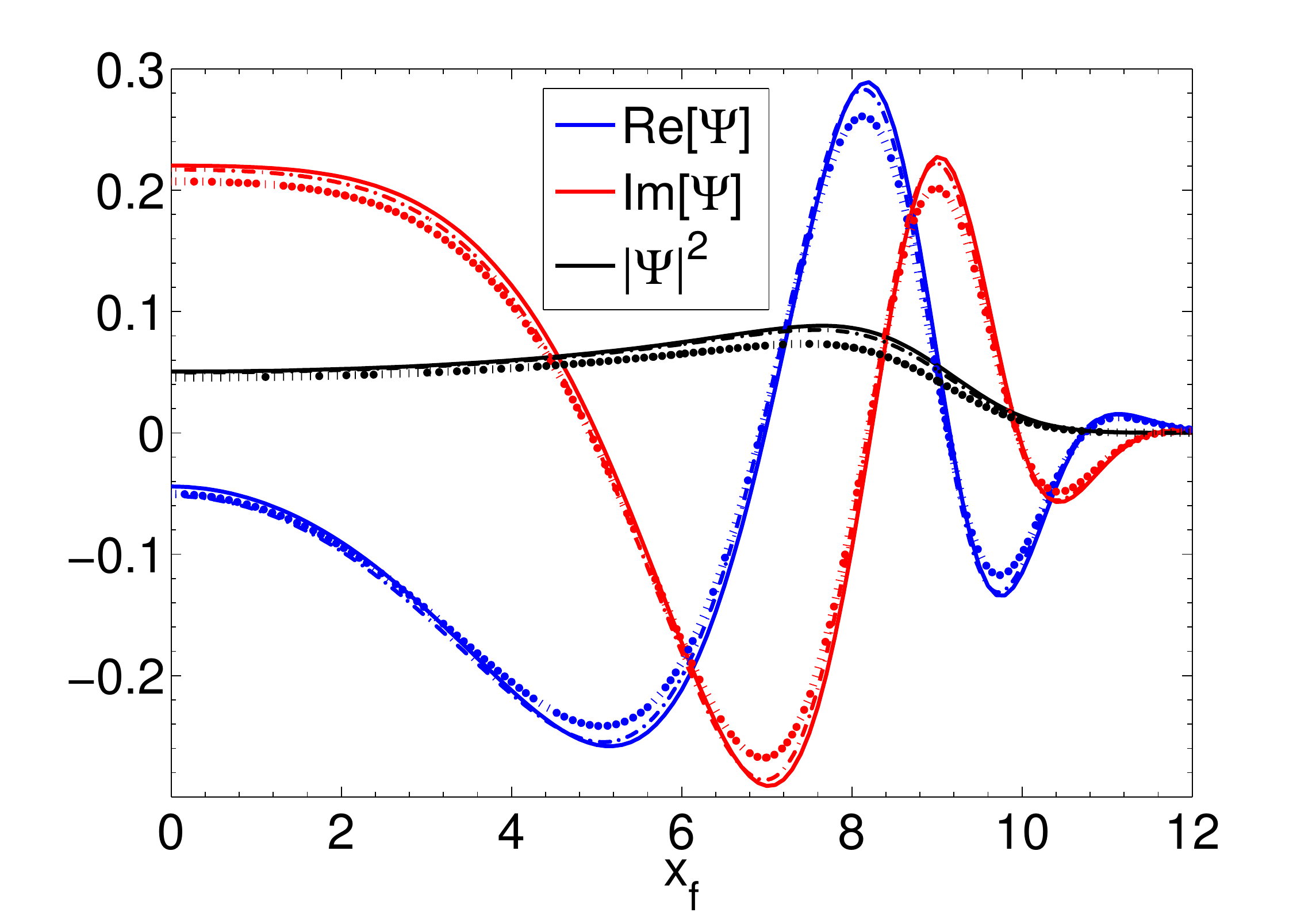}\end{center}
\caption{${\rm Re}[\Psi]$, ${\rm Im}[\Psi]$ and $|\Psi|^2$ are shown for the free particle case at $t=10 \hbar/m$ as a function of $x_f$, which also has 
units of $\hbar/m$. The solid lines show the Schr\"{o}dinger solution, while dotted lines and dashed lines show the WKB approximation for $\epsilon = 0.05$,  $\epsilon = 0.005$ respectively. It can be seen that as $\epsilon \to 0$ 
 the WKB approximation converges to the Schr\"{o}dinger solution.}
\label{freeparticle}
\end{figure}

The WKB propagator for the relativistic free particle is then computed using Eq. (\ref{GWKB}) \cite{1993IJMPA...8.1629R}
\begin{equation}
G_{WKB} =  \sqrt{\frac{m}{2 \pi} \frac{(i \Delta t + \epsilon)^2}{\hbar \lambda_\epsilon^{3/2}}} \exp\left[-\frac{m \lambda_\epsilon^{1/2}}{\hbar}\right],
\end{equation}
where WKB wavefunction is given by
\begin{equation}
\Psi_{WKB}(x,t) = \int_{-\infty}^\infty dx_0 G_{WKB} (x,t; x_0, t_0) \Psi(x_0,t_0).
\label{PsiWKBgen}
\end{equation}
The integrals from Eq.\ (\ref{PsiExact}) and Eq.\ (\ref{PsiWKBgen}) are evaluated numerically. The results are shown in Fig.\ \ref{freeparticle}.
It can be seen that the wavefunction in the WKB approximation converges to the Schr\"{o}dinger solution
in all parts of the light cone. When no star $M$ is present, the relativistic Schr\"{o}dinger
is exact as originally discussed in Redmount and Suen \cite{1993IJMPA...8.1629R}. However, when they 
performed the same numerical comparison, they found some disagreement between the two solutions. This we believe to be due to numerical error of \cite{1993IJMPA...8.1629R}. 
We observe convergence (see Fig.\ \ref{freeparticle}.) of the wavefunction in the WKB approximation to the Schr\"{o}dinger solution in the $M \to 0$ limit.

\begin{figure}\begin{center}
\includegraphics[width=.85\linewidth]{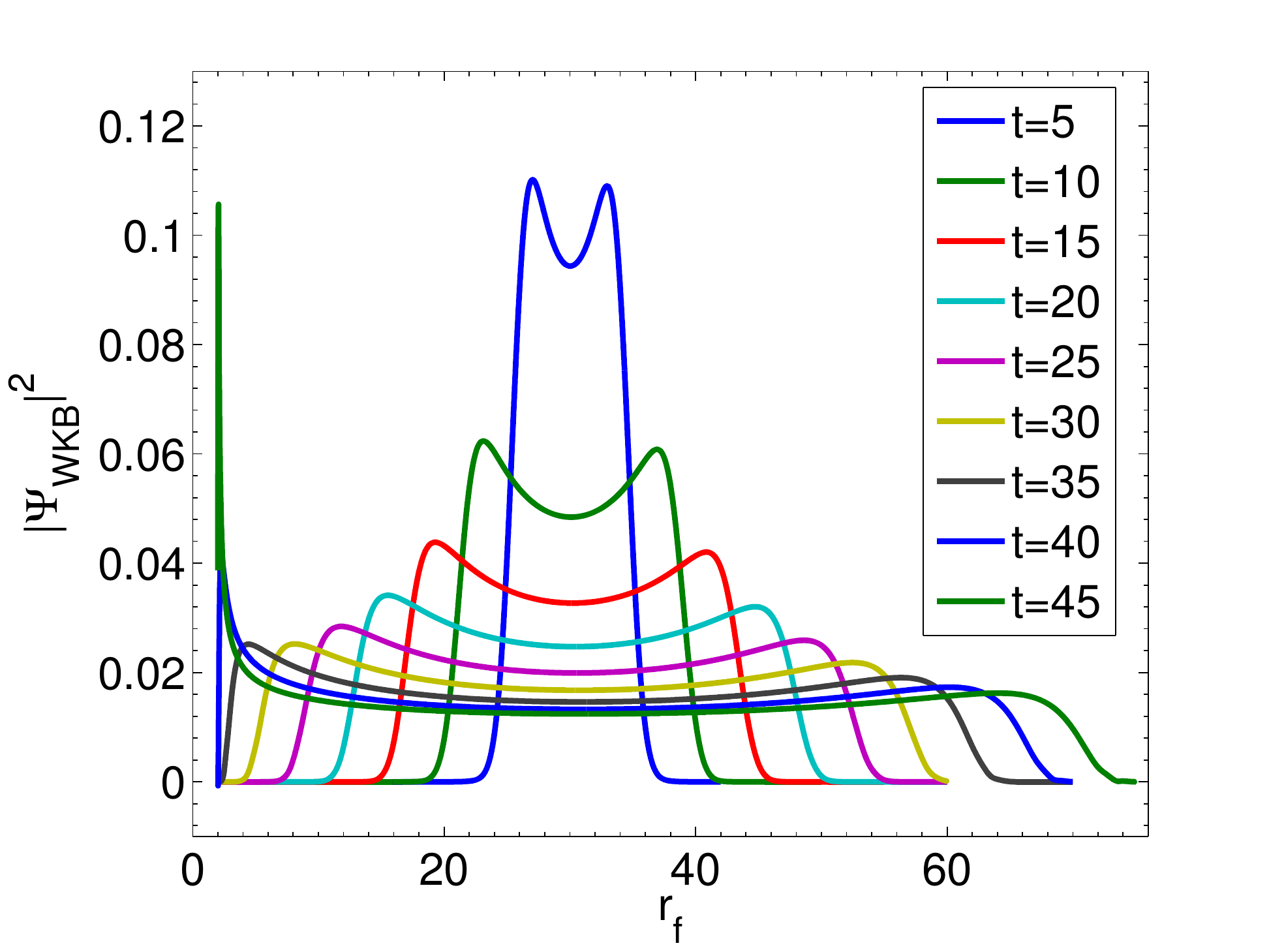} \end{center}
 \caption{\label{Psim1} The time evolution of WKB $|\Psi(r_f, t_f)|^2$ is shown from $t=5 M$ to $t=45 M$ in the case of $m M/\hbar =1$. We can see that the star collapses to a black hole. }
\end{figure}

 \subsubsection{Non-zero mass}
While in the free particle case the Schr\"{o}dinger solution was exact, when $M \ne 0$ it becomes a rough approximation, which fails as we approach $r=2M$. Comparing the two
solutions is still instructive. If we follow the same procedure as for the free particle above, and re-write the  Hamiltonian from Eq. (\ref{Hameq})  as
\begin{equation}
H = \sqrt{B^2 {\bf p}^2 + B m^2}, 
\end{equation}
where ${\bf p}$ is the momentum operator and 
\begin{eqnarray}
B = B(M, r) = 1 - \frac{2 M}{r}.
 \end{eqnarray}
The whole wavefunction can then be written as
\begin{equation}
\Psi(r,t) = \int d r_i G(r, t; r_i, t_i) \Psi(r_i, \tau_i),
\end{equation}
where the propagator is given by
\begin{eqnarray}
G(r, t; r_i, t_i)&=&  \int_{-\infty}^{\infty} \frac{dk}{2 \pi} \exp(i k \Delta r) \\ \nonumber 
&\times& \exp\left[- i \frac{\Delta t}{\hbar} \sqrt{\hbar^2 k^2 B^2 + m^2 B}\right].
\end{eqnarray}
Like before, the propagator can then be integrated exactly to obtain
\begin{eqnarray}
G(x, t; x_i, t_i) = \lim_{\epsilon\to 0} \frac{m (i \Delta t + \epsilon)}{\pi \hbar B^{1/2} \lambda_\epsilon^{1/2}} K_1(m B^{1/2}\lambda_\epsilon^{1/2}/ \hbar),
\label{propSchro}
\end{eqnarray}
where $K_1$ is the modified Bessel function and 
\begin{equation}
\lambda_\epsilon = \left(\frac{\Delta r}{B}\right)^2 + (i \Delta t + \epsilon)^2.
\end{equation}
When $r \to 2 M$ the propagator vanishes. Thus this solution is inaccurate at late times and cannot model the final stages of the 
collapse of the dust sphere.

\begin{figure}\begin{center}
\includegraphics[width=.9\linewidth]{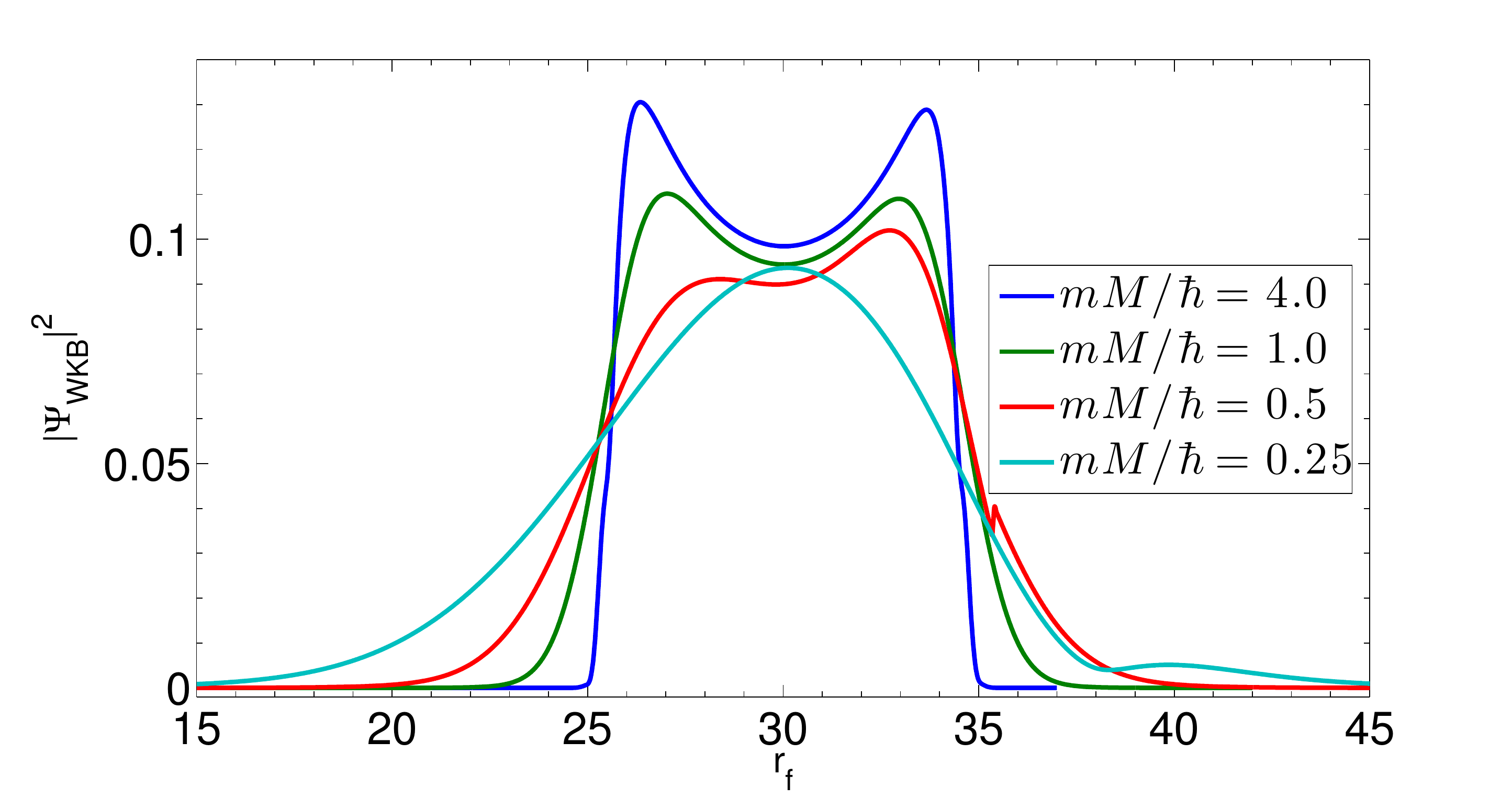}
\includegraphics[width=.85\linewidth]{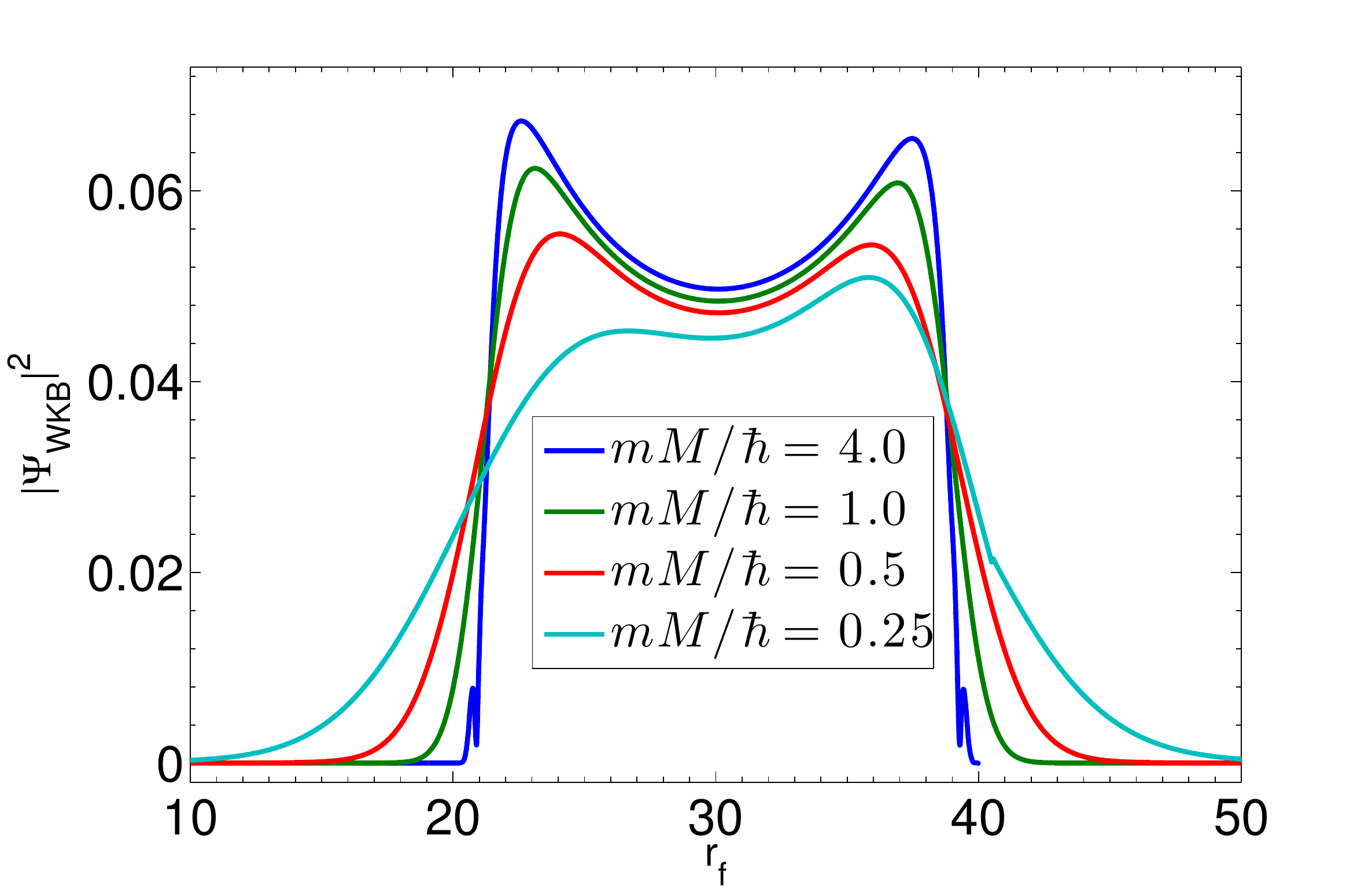}
\end{center}
\caption{\label{probfig} 
(a) $|\Psi(r_f,t_f)|^2$ at $t=5M$ as a function of $r_f$ for $m M/\hbar=4.0$, $m M/\hbar=1$, $m M/\hbar=0.5$ and $m M/\hbar=0.25$. It can be seen that that lower masses behave more quantum
mechanically, and are less likely to collapse to a black hole.
(b) $|\Psi(r_f,t_f)|^2$ at $t=10M$ as a function of $r_f$ for the same masses.}
\end{figure}
\subsection{Numerical Results}
\begin{figure}\begin{center}
\includegraphics[width=.85\linewidth]{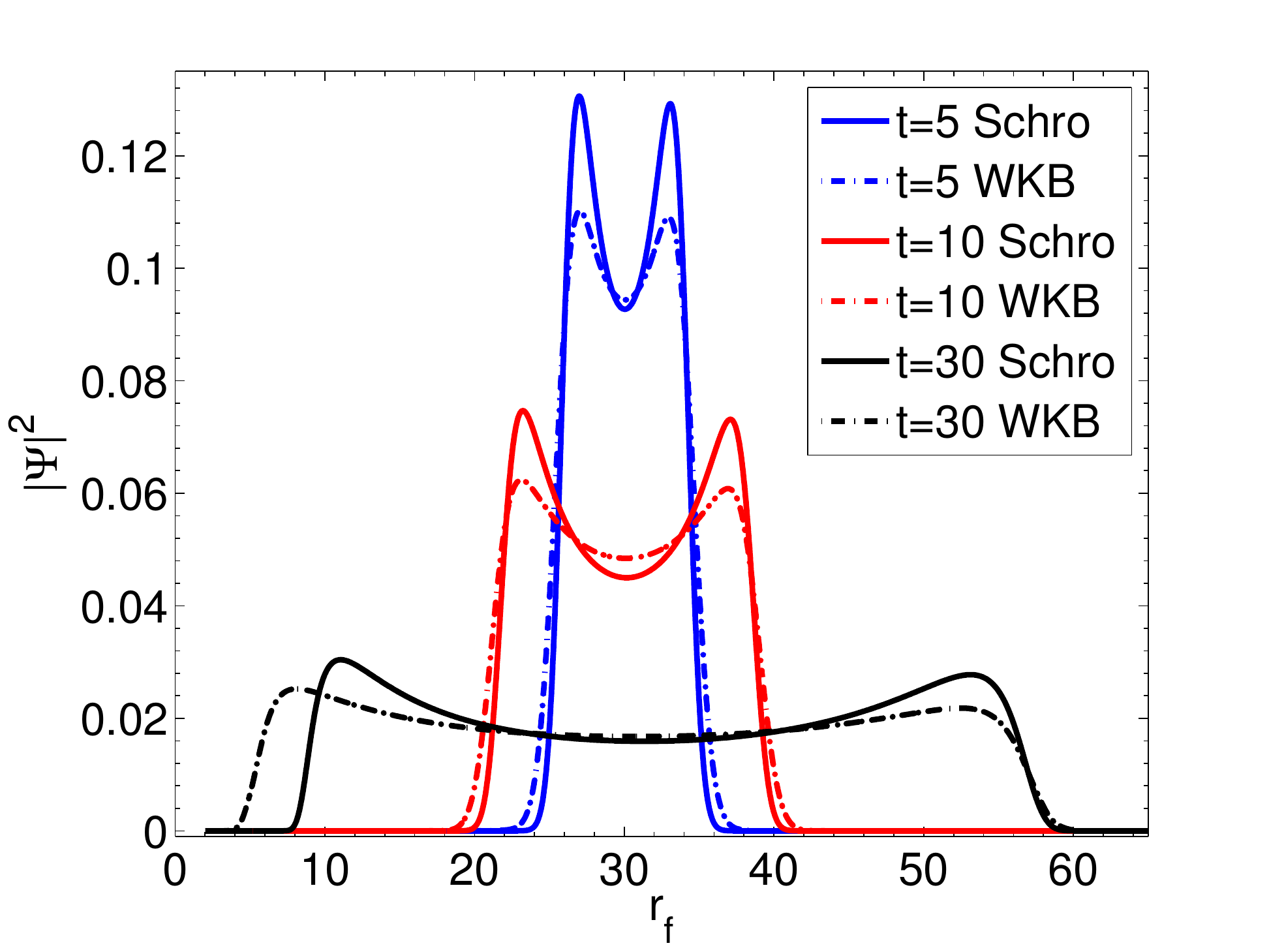} \end{center}
  \caption{\label{SchroWKBcomp1} $m M/\hbar =1$ Schr\"{o}dinger (solid line) and WKB comparison (dotted line). }
\end{figure}
We integrate Eq. \ (\ref{WKBPsi}) numerically for the Gaussian wavefunction from Eq. \ (\ref{Psi0}) centered about $r_c = 30 M$. Classically, the star collapses to $r=2 M$ in infinite Schwarzschild time. However, quantum mechanically there is a possibility for the star expanding and dispersing as well. Fig. \ref{Psim1} shows the time evolution from $t=0$ to $t=45 M$ of $|\Psi_{WKB}(r_f, t_f)|^2$ as a function of $r_f$.  An ingoing peak and an outgoing 
peak appear with the amplitude of the ingoing peak always remaining larger than that of the outgoing peak for the course of the evolution, which is expected for a collapsing star.
For numerical computations we scale all variables to be dimensionless. The only parameter is $m M/\hbar$. We note that Fig. \ref{Psim1} is qualitatively similar to the WKB results of \cite{corichi2002quantum}.

Fig. \ \ref{probfig}(a) and Fig. \ref{probfig}(b) show the ingoing and outgoing peaks of $|\Psi_{WKB}(r, t_f)|^2$ for different values of $m M/\hbar$ at times of $t_f = 5 M$ and $t_f = 10 M$, respectively. For a given particle number $N=M/m$, the speed of the collapse increases slightly with the particle mass $m$ as does the value of the ingoing peak. This is consistent with our expectation that for higher particle mass the star is more classical. For lower masses, the star resists collapse longer until probability of dispersion exceeds the probability of collapse. 

In Fig. \ref{SchroWKBcomp1} we compare the WKB and Schr\"{o}dinger solutions. It can be seen that the radial position of the 
ingoing peak drifts out of step at late times, while the outgoing peak continues to evolve at about the same position. As the star approaches $r_f = 2 M$, the ingoing peak amplitude for the WKB 
approximation increases developing a numerical singularity that indicates the formation of an apparent horizon,  
while the ingoing peak amplitude of the Schr\"{o}dinger solution decreases since the propagator in Eq.\ (\ref{propSchro}) vanishes at $r_f \to 2 M$. 
The latter behaviour reinforces that the relativistic Schr\"{o}dinger equation is not the correct
representation for a particle in the gravitational field of a collapsing star.

\section{Conclusion}
\label{sec:conclusions}
Closed form solutions for the classical paths taken by a particle on the surface of a collapsing star 
have been determined for all initial configurations
in Schwarzschild 
and Kruskal coordinates.

We found that
\begin{itemize}
\item[(i)] all time-like paths are unique. For a given time interval,  a path between an initial and final point can be either direct or indirect, where it turns around in space. Thus some particles that initially move away from the star, can return and contribute to the collapse.
\item[(ii)] all space-like paths are unique outside $r=2M$.
\item[(iii)]  Kruskal space-like paths can turn around in Kruskal time, but cannot turn in Kruskal space. Multiple paths can connect a given point
outside the horizon to a final point that lies inside $r=2 M$.  
\item[(iv)] 
Classical Kruskal space-like paths connect points outside $r = 2 M$ with points inside $r=2M$ up to a critical value of $r= r_a$. Spacelike paths from ($r_i, v_i=0$) to any given point 
$r$ inside the horizon, where $r_a <r< 2M$, are non-unique  (two paths exist). These paths get less and less separated and merge into a single space-like path at $r=r_a$ and any point $r<r_a$ is unreachable from $r_i$.
Upon reaching the critical value $r= r_a$, the space-like paths turn back towards $r=2M$, reaching it at $u=v=0$. Therefore, classically, no information from $r < 2 M$ can exit a black hole. However, by taking into account paths close to the classical paths, one might be able to extract information from $r < 2M$.

If we make an analogy to tunneling, the particle has a low probability
of passing through a potential barier. Classically, it never happens, and yet this probability cannot be ignored 
in quantum mechanics. Similarly, it may be that the low probability space-like paths will play a crucial role in
our understanding of quatum mechanical collapse.
\end{itemize}

The collapse of a self-gravitating star was next modeled as a ball of dust using a WKB approximation. 
We integrated around the classical paths with all possible initial velocities to include quantum effects.
This extended the analysis of  \cite{1993IJMPA...8.1629R} 
from the 
relativistic free particle to the case of non-trivial gravity. In practice, quantum mechanical effects are important 
in macroscopic stellar collapse when the sphere is composed of ultra-light particles. 
A number of ultralight dark matter particles have been proposed, which could physically motivate such stars.  

The evolution of the wavefunction of a particle on the surface of a collapsing star was followed numerically.
We showed that in the case of a star collapsing with zero initial velocity, our path equations 
reduce to the Oppenheimer-Schneider equations of motion. 
In the $M \to 0$ limit, the relativistic free particle wavefunction of \cite{1993IJMPA...8.1629R} is obtained, and
the WKB and relativistic Schr\"{o}dinger solutions match. Since in this limit 
the Schr\"{o}dinger solution  is exact \cite{kiefer1991quantum}, the covergence of the WKB to the Schr\"{o}dinger solution 
enforces the validity of the WKB approximation.

Our results for this self-gravitating collapse are summarized as follows:
\begin{itemize}
\item[(i)] 
The wavefunction representing a particle on the surface of a collapsing star typically exhibits an outgoing and an ingoing component, where the former contains the probability that the star will disperse and the latter the collapse probability. For a given particle number $N=M/m$, we find that the rate at which collapse occurs increases with particle mass. This is consistent with the expectation that for higher particle mass the star is more classical.  As the particle mass is lowered, the star resists collapse until the probability that it disperses exceeds the collapse probability.
Note that some part of the star always disperses even when a black hole forms.
 
\item[(ii)]  In the case of the collapsing star the relativistic Schr\"{o}dinger solution is not a good approximation for the wave function.
On comparing the WKB and relativistic Schr\"{o}dinger solutions, we find that the ingoing wavefunction gets more and more out of step
at late times. The outgoing component of the wavefunction for the two solutions is in better agreement because it is further from the coordinate singularity at $r=2 M$. 
As expected, the Schr\"{o}dinger and WKB solutions are out of step at early times when WKB solution is more inaccurate, they come closer together at intermediate times, and fall out of step again at late times, when the Schr\"{o}dinger approximation fails to model the singularity formation.  
The presence of the coordinate singularity at $r=2 M$, motivate a potential investigation beyond $r=2M$ via the WKB approximation deployed in Kruskal coordinates where no analytical relativistic Schr\"{o}dinger solution is available. 
\end{itemize}

		The traditional Oppenheimer Schneider model describes the zero pressure collapse of a star starting from rest. We use a path integral approach that requires including paths of all velocities. For a particle on the surface of a star, we explicitly parametrically describe all such paths of different initial velocities.  This is done in Schwarzschild coordinates until $r=2M$ and in Kruskal coordinates until the $r=0$ singularity. We include ingoing and outgoing velocities, and time-like, light-like and space-like paths. In the process we have generated the paths shown in \cite{fuller1962causality} and more. We have determined equations for the space-like caustics (degenerate paths) which have not been described before in this manner. We have used a WKB path integral approach and determined the Schwarzschild wave function for the collapse. Using a similar approach one can in the future use the parametric path equations we have determined in Kruskal coordinates to explore the space-time inside the horizon. It will require expansion around the caustic. The possibility of paths moving back in time might have important ramifications.

It has long been argued whether quantum gravity should be causal or not \cite{teitelboim1983causality,hartle1988quantum,hartle1986path,lrr-2005-5} and as of today there is no accepted theory of quantum gravity.
We evaluate the propagator both for a free relativistic particle, where we have an exact solution, and for a collapsing ball of dust. 
Since the propagator does not vanish outside the light cone, we include all paths in our integration. In the free particle case, the WKB 
and exact solutions match. Space-like classical paths give the dominant contribution in the construction of the WKB approximation far 
outside the light cone, where the approximation is valid. This propagation will be acausal (backward in time) in some Lorentz frames. 
 Whether or not this proves to be admissible in the ultimate theory of quantum gravity is beyond the purpose of our work. Our work's 
exploration of the consequence of an acausal propogator in the collapse of a zero pressure star might nevertheless contribute to the dialogue in the search of a consistent theory of quantum gravity. 

In addition, we propose that an understanding of these low probability paths that are traditionally ignored could shed some light in theories of black hole formation. Furthermore, if the black holes observed by LIGO are formed from scalar particles \cite{abbott2016astrophysical}, the particles would be very light; then quantum effects might play
a role in their formation, constraining analogue gravity theorems \cite{barcelo2005analogue}.

\section*{Acknowledgments}
JB acknowledges Prof.\ Wai-Mo Suen for the initial impetus to approach this research and for subsequent guidance and 
useful discussions. We also thank Prof. Ian Redmount for his inital work on this topic. Further, we are particularly grateful to Prof. Mihai Bondarescu and Prof.\ Philippe Jetzer for useful discussions and advice.  RB has received support from the Dr.\ Tomalla Foundation and the Swiss National Science Foundation. CCM  is supported by the NSF Astronomy and Astrophysics Postdoctoral Fellowship under award AST-1501208.



\bibliography{ms}
\end{document}